\documentclass[apj]{emulateapj}
\usepackage{apjfonts,amsmath}
\makeatletter

\newcommand{\Rmnum}[1]{\expandafter\@slowromancap\romannumeral #1@}
\makeatother

\newcommand{\gps}{\ensuremath{g_{\rm P1}}}
\newcommand{\rps}{\ensuremath{r_{\rm P1}}}
\newcommand{\ips}{\ensuremath{i_{\rm P1}}}
\newcommand{\zps}{\ensuremath{z_{\rm P1}}}
\newcommand{\yps}{\ensuremath{y_{\rm P1}}}

\newcommand{\grizy}{\gps\rps\ips\zps\yps}
\newcommand{\griz}{\gps\rps\ips\zps}

\newcommand{\PS}{\protect \hbox {Pan-STARRS1}}

\slugcomment{Unpublished Draft}

\shorttitle{PS1 SDSS Quasar Variability}
\shortauthors{Morganson et al.}

\begin{document}


\title{Measuring Quasar Variability with \PS and SDSS}


\author{E. Morganson\altaffilmark{1}, W. S. Burgett\altaffilmark{2}, K. C. Chambers\altaffilmark{2}, P. J. Green\altaffilmark{3},  N. Kaiser\altaffilmark{2}, E. A. Magnier\altaffilmark{2}, P. J. Marshall\altaffilmark{4}, J. S. Morgan\altaffilmark{2}, P. A. Price\altaffilmark{5}, H.-W. Rix\altaffilmark{1}, E. F. Schlafly\altaffilmark{1}, J.L. Tonry\altaffilmark{2}, F. Walter\altaffilmark{1} }
\email{morganson@mpia.de}


\altaffiltext{1}{Max-Planck-Institut f\"ur Astronomie, K\"onigstuhl 17, 69117 Heidelberg, Germany}
\altaffiltext{2}{Institute for Astronomy, University of Hawaii at Manoa, Honolulu, HI 96822, USA}
\altaffiltext{3}{Harvard-Smithsonian Center for Astrophysics, 60 Garden Street, Cambridge, MA 02138, USA}
\altaffiltext{4}{Department of Astrophysics, Oxford University, Denys Wilkinson Building, Kemble Road, Oxford, OX1 3RH, UK}
\altaffiltext{5}{Department of Astrophysical Sciences, Princeton University, Princeton, NJ 08544, USA}


\begin{abstract} 
We measure quasar variability using the Panoramic Survey Telescope and Rapid Response System 1 Survey (\PS\ or PS1) and the Sloan Digital Sky Survey (SDSS) and establish a method of selecting quasars via their variability in $10^4$ square degree surveys. We use $10^5$ spectroscopically confirmed quasars that have been well measured in both PS1 and SDSS and take advantage of the decadal time scales that separate SDSS measurements and PS1 measurements. A power law model fits the data well over the entire time range tested, 0.01 to 10 years. Variability in the current PS1-SDSS dataset can efficiently distinguish between quasars and non-varying objects. It improves the purity of a $griz$ quasar color cut from 4.1\% to 48\% while maintaining 67\% completeness. Variability will be very effective at finding quasars in datasets with no u band and in redshift ranges where exclusively photometric selection is not efficient. We show that quasars' rest-frame ensemble variability, measured as a root mean squared in $\Delta$magnitudes, is consistent with $\rm{V}(\rm{z},\rm{L},\rm{t}) = \rm{A}_0 (1+\rm{z})^{0.37}(\rm{L}/\rm{L}_0)^{-0.16} (\rm{t}/1 yr)^{0.246}$, where $\rm{L}_0 = 10^{46} \rm{erg} \rm{s}^{-1}$ and A$_0$ = 0.190, 0.162, 0.147 or 0.141 in the \gps, \rps, \ips or \zps filter, respectively. We also fit across all four filters and obtain median variability as a function of z, L and $\lambda$ as $\rm{V}(\rm{z},\rm{L},\lambda,\rm{t}) = 0.079 (1+\rm{z})^{0.15}(\rm{L}/\rm{L}_0)^{-0.2} (\lambda/1000\ nm)^{-0.44} (\rm{t}/1 yr)^{0.246}$.

\end{abstract}


\keywords{(Galaxies:) quasars: general  }

%% In the first two sections, notice the use of the natbib () or \object{}.  Each macro takes the

\section{Introduction}\label{intro}

Quasars are supermassive black holes in the centers of galaxies that have large accretion rates and correspondingly large luminosities ({Rees} 1984; {Antonucci} 1993; {Kembhavi} \& {Narlikar} 1999). They can be up to one hundred times brighter than their host galaxies (e.g. {Villata} {et~al.} 2006) and can thus be observed and analyzed in depth even when their host galaxies are not observable. Because of this, quasars up to $z = 7.1$ have been detected and spectroscopically analyzed ({Fan} {et~al.} 2001; {Mortlock} {et~al.} 2011; {Morganson} {et~al.} 2012), and moderately high redshift ($z > 2$) lensed quasars are among the most commonly detected galaxy scale gravitational lenses ({Oguri} {et~al.} 2006; {Inada} {et~al.} 2008; {Oguri} \& {Marshall} 2010). 
%Quasar lenses are in turn used to understand the molecular gas of distant galaxies ({Yun} {et~al.} 1997; {Riechers} {et~al.} 2007a, 2007b), dark matter structure ({Dalal} \& {Kochanek} 2002; {Brada{\v c}} {et~al.} 2002; {Dobler} \& {Keeton} 2006) and other phenomena.

\begin{figure}[ht]
\includegraphics[width=0.99\columnwidth]{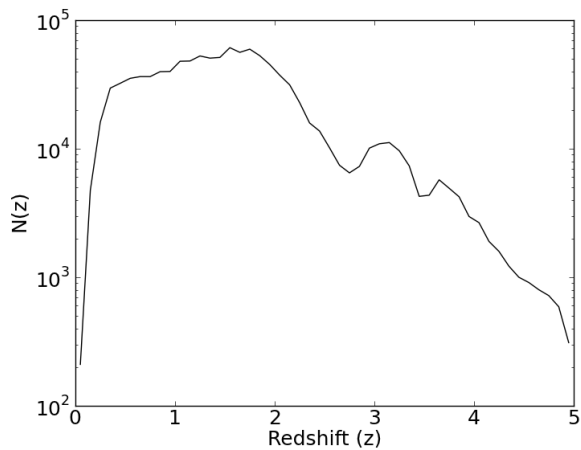}
\caption{The number per unit redshift of spectroscopically confirmed quasars from {Shen} {et~al.} (2011). The lack of u dropouts at z > 2 as well as the g dropout quasar population at z $\approx$ 3.5 are both evident. }
\label{fig:f1}\end{figure}

Despite the huge efforts undertaken to find quasars, there is a dearth of known quasars at $z > 2.5$ ({Schneider} {et~al.} 2002, 2010) as shown in Fig. \ref{fig:f1}. Survey depth is the major reason for this, but quasar selection incompleteness also plays a significant role. Nearly all quasars were initially detected in large photometric surveys like the Sloan Digital Sky Survey (SDSS, {York} {et~al.} 2000). In these surveys, quasars are primarily detected as objects with excess $u'$ band flux ({Richards} {et~al.} 2002, 2004). Quasars have relatively flat continuum spectra with no thermal dropoff and thus tend to be blue and particularly $u'$-bright relative to nearly any star or non-quasar galaxy. At $z > 2$, the quasars' rest-frame Ly-$\alpha$ absorption enters the observer-frame $u'$ band, and quasars cease to have exceptional $u'$ band fluxes. For $z > 2.5$ quasars, $u'$ band excess selection is no longer viable. This Ly-$\alpha$ absorption becomes useful for higher redshift quasars, because the sudden dropoff in flux is also photometrically distinctive ({Fan} 1999). But high redshift "dropout" searches only allow us to find quasars in relatively small redshift ranges where the dropout effect is distinct ({Osmer} 1982; {Warren} {et~al.} 1991). In SDSS, $z \approx 3.5$ quasars are $g'$ dropouts, $z \approx 4.5$ quasars are $r'$ dropouts, and $z \approx 6$ quasars are $i'$ dropouts. The result is that the population of $z > 2.5$ quasars is relatively small and nonuniform in redshift space.

Quasars vary nonperiodically in optical bands by several tenths of a magnitude over periods of months and years ({Giveon} {et~al.} 1999; {Vanden Berk} {et~al.} 2004). The main causes of this variability include accretion disk instabilities ({Rees} 1984; {Kawaguchi} {et~al.} 1998; {Pereyra} {et~al.} 2006) and inflow variation ({Hopkins} {et~al.} 2006). Microlensing by intervening lensing galaxies ({Wambsganss} 2006; {Morgan} {et~al.} 2010) also contributes in some cases. Regardless of which physical processes are responsible, optical quasar variability is distinctive and can be used to more efficiently select quasars ({Koz{\l}owski} {et~al.} 2010; {Butler} \& {Bloom} 2011; {Palanque-Delabrouille} {et~al.} 2011; {Schmidt} {et~al.} 2010; {MacLeod} {et~al.} 2011) as has already been done in relatively small (fewer than 1000 quasar) surveys ({Geha} {et~al.} 2003; {Kozlowski}, {Kochanek}, \&  {Udalski} 2011; {Kim} {et~al.} 2011; {Koz{\l}owski} {et~al.} 2012, {Koz{\l}owski} {et~al.} 2013). 

Before we can use quasar variability to search for new quasars, we must understand, statistically, how quasars vary. Previous attempts to measure quasar variability have been limited by time range or sample size. {Vanden Berk} {et~al.} (2004) measured the variablity of 25,000 spectroscopically confirmed SDSS quasars by comparing photometry derived from follow-up spectroscopy to the initial detection photometry. This work was limited to a maximum time lag between photometry and spectroscopy of two years. {Schmidt} {et~al.} (2010) and {MacLeod} {et~al.} (2010) studied quasar variability in SDSS stripe 82 ({Abazajian} {et~al.} 2009) on time scales up to 5 years, but could only use the 9,157 quasars in stripe 82. {MacLeod} {et~al.} (2012) extends the SDSS work even further by tracking variability of the 33,881 quasars that were either imaged in stripe 82 or were multiply imaged due to overlaps in the larger SDSS surveys. Exploiting the full sample of the 105,783 spectroscopically confirmed quasars from {Shen} {et~al.} (2011), and the more than 10 years of time lag since SDSS began producing massive samples of quasars would increase our knowledge of quasar variability further. 

The Panoramic Survey Telescope and Rapid Response System 1 (PS1, {Kaiser} {et~al.} 2010) and its $3\pi$ survey are powerful new tools for studying a new statistical regime of quasar variability. 
The PS1 $3\pi$ survey images the entire sky north of -30 degrees declination, including the entire Sloan Digital Sky Survey. It thus produces new photometry of the 10$^5$ spectroscopically confirmed SDSS quasars. The time difference between SDSS and PS1 measurements of an individual quasar are typically 5-10 years. 

For this paper, we cross-matched the PS1 and SDSS databases with the spectroscopically confirmed quasars from {Shen} {et~al.} (2011) to precisely measure quasar variability and show that variablity can be used to find new quasars in the future. We describe the PS1-SDSS cross-matched database that we produced and used for this work in the next section. In sections \ref{sect:var} and  \ref{sect:var2}, we describe how we measure and parameterize variability for every object in this database, including quasars. In section \ref{sect:drw}, we briefly discuss the damped random walk model of quasar variability. In section \ref{sect:varobs}, we analyze quasar variability in the observer-frame. We discuss the average PS1-SDSS magnitude offset for quasars in section \ref{sect:offset}. In section \ref{sect:varsel}, we show that the variability measurements from the cross-matched PS1-SDSS database can be used to improve quasar selection efficiency significantly. In sections \ref{sect:varrest}, we measure the rest-frame variability of the Shen quasars. Finally, we show how variability amplitude relates to luminosity, redshift and wavelength in section \ref{sect:varzl}. 

\section{The PS1-SDSS Dataset}\label{sect:data}

PS1 ({Kaiser} {et~al.} 2002, 2010; {Chambers} 2011) is a 1.8 m optical telescope with a 7 degree$^2$ field of view that images the sky in the $\gps$, $\rps$, $\ips$ and $\zps$ filters which cover the $4000\rm{\AA} < \lambda < 9200\rm{\AA}$ spectral range similarly to the analogously-named SDSS $g'$, $r'$, $i'$ and $z'$ filters. It also has a $\yps$ filter which, including the spectral response of the camera, covers the $9200\rm{\AA} < \lambda < 10500 \rm{\AA}$ range. These filters are described in detail in {Tonry} {et~al.} (2012). The telescope is producing several surveys including a solar system Near Earth Object survey, a Stellar Transit Survey, a Deep Survey of M31, a Medium Deep survey consisting of 10 PS1 footprints spaced around the sky and a $3\pi$ Survey which covers $3/4$ of the sky (30,000 square degrees) in all five bands ({Chambers} 2011). This latter survey is the focus of our work here as it contains the entire SDSS survey and approximately $10^5$ spectroscopically confirmed quasars. 

The PS1 $3\pi$ survey takes four exposures per year with each of the $\grizy$ filters. The yearly fill factor is roughly 90\% in each band. The missing area is due mostly to non-detection areas on the camera plane and weather restricting the survey to 2 or rarely 0 exposures per filter in some areas of the sky. Individual $\grizy$ exposures have median 5$\sigma$ limiting AB magnitudes of 22.1, 21.9, 21.6, 20.9 and 19.9, as summarized in Table \ref{tab:limmag}. These PS1 limits are median results within the SDSS area. Stacked images are not uniformly available, and the work presented here is based on single exposure detections. However, when stacks are made, we expect a single year's stacked image to increase each limiting magnitude by approximately 0.7 (accounting for some survey incompleteness), and the stacks of the proposed three year duration of the survey to increase them by 1.2. In this work, we use the \"ubercalibrated data from {Schlafly} {et~al.} (2012). This database includes 4/5 of the PS1 data up through August 2012 and is calibrated to 0.01 magnitudes or better. The remaining 1/5 of the data was excluded because it could not be calibrated due to weather or technical issues. This database also excludes detections flagged by PS1 as cosmic rays, edge effects and other defects. The PS1 $3\pi$ survey is still collecting more data, and the techniques developed here will become even more powerful as new epochs are added to the database. 

\begin{table}
\begin{tabular}{ccccc}
	\hline
Filter &  SDSS    &   PS1 Exposure & PS1 1 Year & PS1 3 Year  \\
	\hline
u   & 22.3 & --   & --   & --   \\
g   & 23.3 & 22.1 & 22.8 & 23.3 \\
r   & 23.1 & 21.9 & 22.6 & 23.1 \\
i   & 22.3 & 21.6 & 22.3 & 22.8 \\
z   & 20.8 & 20.9 & 21.6 & 22.1 \\
y   & --   & 19.9 & 20.6 & 21.1 \\
	\hline
\end{tabular}
\caption{\rm{5$\sigma$ Limiting AB Magnitudes of SDSS ({York} {et~al.} 2000) and PS1 3$\pi$. 1 Year and 3 Year stack results are predicted. Similarly-named filters from different surveys are not exactly the same.}}\label{tab:limmag}
\end{table}

We use SDSS photometry from SDSS Data Release 8 ({Aihara} {et~al.} 2011). SDSS DR8 covers 14,555 square degrees to $u'g'r'i'z'$ to 22.3, 23.3, 23.1, 22.3 and 20.8. To convert between SDSS and PS1 magnitudes, we use the conversions from Finkbeiner et al. (2013, in preparation) which use the equation
\begin{eqnarray} 
m_{\rm{P1}}-m_{\rm{SDSS}} &=& \rm{a}_0 + \rm{a}_1\ gi + \rm{a}_2\ gi^2 + \rm{a}_3\ gi^3,\nonumber\\ 
gi &=& g_{\rm{SDSS}}-i_{\rm{SDSS}}\label{eq:spcon}
\end{eqnarray}
where $m = griz$ and a$_{0123}$ are in Table \ref{tab:coeff}. {Tonry} {et~al.} (2012) provides a similar conversion from SDSS to PS1 calculated from PS1 filter curves. But we use the Finkbeiner conversion because it is optimized for a broad stellar population. It is also calculated within the {Schlafly} \& {Finkbeiner} 2012 \"ubercalibrated system which we are using for our photometry. For the non-varying stars for which these coefficients were fit, these conversions are good to roughly 0.01 mag. We add this 0.01 mag in quadrature to our statistical error. In Section \ref{sect:varrest} we find that these conversions are not as accurate for quasars and that we must use an additional correction to match PS1 to SDSS. When using SDSS magnitudes, we convert them to standard logarithmic magnitudes, rather than the default arcsinh-based "Luptitudes" that SDSS reports.

\begin{table}
\begin{tabular}{ccccc}
	\hline
Filter &  a$_0$   & a$_1$    & a$_2$    & a$_3$    \\
	\hline
g      & 0.00128  & -0.10699 & 0.00392  & 0.00152  \\
r      & -0.00518 & -0.03561 & 0.02359  & -0.00447 \\
i      & 0.00585  & -0.01287 & 0.00707  & -0.00178 \\
z      & 0.00144  & 0.07379  & -0.03366 & 0.00765  \\
	\hline
\end{tabular}
\caption{\rm{The coefficients used to convert from SDSS magnitude to PS1 magnitudes in Eq. \ref{eq:spcon}. Ensemble error bars are insignificant, and for individual stars, these conversions are good to 0.01 magnitudes.}}\label{tab:coeff}
\end{table}

\begin{figure}[ht]
\includegraphics[width=0.99\columnwidth]{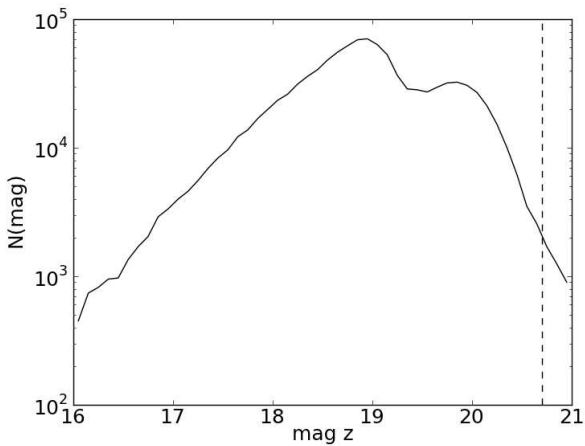}
\caption{\rm{The number per unit magnitude of spectroscopically confirmed quasars from {Shen} {et~al.} (2011). The vast majority are brighter than the median PS1 limit of \zps = 20.7 as shown by the dotted line. We have converted SDSS z' magnitudes into the PS1 photometric system .} } 
\label{fig:f2}\end{figure}

We use the 105,783 spectroscopically confirmed quasars from {Shen} {et~al.} (2011) as our known quasar population. These quasars all have redshifts and model bolometric luminosities. They are quite bright, as we see in the z band distribution in Fig. \ref{fig:f2}, and 98.7\% have PS1-converted SDSS magnitudes brighter than the PS1 limiting magnitude in each of the four PS1 bands in Table \ref{tab:limmag}. We considered analyzing  the sample of $\approx 10^6$ SDSS photometric quasars ({Richards} {et~al.} 2009). But the sources in that sample are not all quasars, do not have precisely measured redshifts or luminosities, and many would be too faint for single epoch PS1 detections. 

All database work and cross-matching of surveys is performed with the Large Survey Database software ({Juric} 2011). We cross-match and parameterize the variability of all 252,567,124 objects that exist in both PS1 and SDSS DR8. We make a separate catalog for the spectroscopic quasars from {Shen} {et~al.} (2011). We compare PS1 and SDSS PSF magnitudes in all cases and take care to exclude extended objects (with techniques described in section \ref{sect:varobs}) from our analysis in this paper. 

\section{The Structure Function}\label{sect:var}

To parameterize quasar variability efficiently in large surveys, we modify the quasar variability statistic from {Schmidt} {et~al.} (2010). This statistic assumes that for an individual quasar in an individual filter, the magnitude difference, m, between two measurements, j and k, separated by a time difference t $=$ t$_j$-t$_k$ is a Gaussian with the form
\begin{eqnarray}
\rm{P}(\rm{m}) &=& \frac{1}{\sqrt{2\pi (\rm{V}^2(\rm{t})+\sigma^2)}}e^{-\frac{\rm{m}^2}{2(\rm{V}^2(\rm{t})+\sigma^2)}},\nonumber\\ 
\rm{V}^2(\rm{t} \mid \rm{A}, \Gamma)&=& \rm{A}^2 \left(\frac{t}{1\ \rm{yr}}\right)^\Gamma,\nonumber\\ 
\sigma^2&=&\sigma_j^2+\sigma_k^2.\label{eq:v}
\end{eqnarray}
Here, P(m) is the normalized probability distribution of m, and $\sigma_j^2$ and $\sigma_k^2$ are the statistical magnitude uncertainties for quasar measurements j and k so that $\sigma^2$ is the statistical uncertainty of the magnitude difference m. {MacLeod} {et~al.} (2012) finds that ensemble quasar variability is consistent with a Gaussian model for individual quasar variability. V(t), known as the structure function, is an ensemble measurement and fit of quasar variability. Formally, V$^2$(t) is the average variance of m due to actual astrophysical variation, and V(t) is just the root mean squared of the average physical variability in units of magnitudes. The power law V$^2(\rm{t}) = \rm{A}^2 \rm{t}^\Gamma$ is just an empirical fit and not a physical model. In this paper, the time difference t is always in years (observer-frame in \ref{sect:varobs} and rest-frame in section \ref{sect:varrest}). A is then just the amplitude of variation at one year. While V(t) = A t$^\gamma$ is the more intuitive function, for mathematical ease, we use V$^2$, A$^2$ and $\Gamma = 2\gamma$ in our derivations and convert into the conventional A and $\gamma$ for plots and final results.

%Other studies ({Kelly}, {Bechtold}, \&  {Siemiginowska} 2009; {Koz{\l}owski} {et~al.} 2010; {MacLeod} {et~al.} 2010) found that individual quasars are well modeled as damped random walks with each quasar getting a characteristic time scale. Our results are essentially consistent with a power law, and when measuring variability for $10^9$ sources across five filters, the power law provides a major computational advantage. In section \ref{sect:drw} we find that an analytic approximation of an ensemble damped random walk model describes our ensemble observations slightly better than a power law model. 

\section{Parameterizing Quasar Variability}\label{sect:var2}

To apply Eq. \ref{eq:v} to a practical problem, we start with a quasar that has been measured a total of N times in a PS1 or (PS1-converted) SDSS filter. A series of N measurements produces n = N(N-1)/2 discrete magnitude differences. We use $i$ to index over n. Following the normal $\chi^2$ type derivation, we find that the $\log$ probability of Eq. \ref{eq:v} is
\begin{eqnarray}
\log \rm{P}(\rm{m_i}, \sigma_i, \rm{t}_i \mid \rm{A}^2, \Gamma) &=& -\frac{1}{2} \sum_i \log 2\pi +\log(\rm{V}^2(\rm{t_i})+\sigma_i^2)\nonumber\\
 + \frac{\rm{m}_i^2}{\rm{V}^2(\rm{t}_i)+\sigma_i^2}.\label{eq:logp}
\end{eqnarray}

{Schmidt} {et~al.} (2010) maximizes Eq. \ref{eq:logp} using Monte Carlo methods. Since we are implementing this over $10^9$ sources, we search for faster, analytic approximations. We assume that $\rm{V}^2(\rm{t}) = \rm{A}^2 \rm{t}^\Gamma$, and in our derivations, we fix $\Gamma$, taking it from the fitted ensemble value that we describe at the end of this section. Even with this assumption, it is difficult to analytically maximize Eq. \ref{eq:logp} for an arbitrary set of t$_i$'s. To simplify the problem further, we bin our t$_i$'s so that within a bin the time lags are roughly constant. We then analytically maximize the likelihood in Eq. \ref{eq:logp} of each bin with respect to A$^2$ and include the differences in time lags as a first order perturbation. We perform a lengthy derivation in Appendix A. The final result is that our estimator for A$^2$ for each quasar in each time lag bin is 
\begin{eqnarray}
\rm{A}^2&=&\frac{2\rm{m}_0^2-\sigma_0^2}{\rm{t}_0^\Gamma}-\frac{1}{\rm{n}\ \rm{t}_0^\Gamma}\sum_i \frac{\rm{m}_i^2 \rm{t}_i^\Gamma}{\rm{t}_0^\Gamma}\nonumber\\ 
&&+\frac{2}{\rm{n}\ \rm{t}_0^\Gamma\left(\rm{A}^2_{\rm{model}}\rm{t}_0^\Gamma+\sigma_0^2\right)}\left(\sigma_0^2 \sum_i \frac{\rm{m}_i^2 \rm{t}_i^\Gamma}{\rm{t}_0^\Gamma}-\sum_i \rm{m}_i^2 \sigma^2_i\right),\nonumber\\
\rm{m}_0^2&=&\frac{1}{\rm{n}}\sum_i \rm{m}_i^2,\ \sigma^2_0=\frac{1}{\rm{n}}\sum_i \sigma_i^2,\ \rm{t}_0^\Gamma=\frac{1}{\rm{n}}\sum_i \rm{t}_i^\Gamma\label{eq:a2}. 
\end{eqnarray}
Here $i$ iterates over measurement pairs with time lags in this bin. A$_{\rm{model}}$ is a model value for A which we take as the ensemble fit value that we describe at the end of this section. In practice, small variations in A$_{\rm{model}}$ do not tend to affect the estimate of A very much, because the terms in Eq. \ref{eq:a2} for which A$^2_{\rm{model}}$ is in the denominator average to zero. 

In Appendix A, we also derive an inverse variance weight
\begin{eqnarray}
\rm{weight}&=&\frac{\rm{\hat{n}}}{2\rm{n}} \sum_i \frac{1}{\left(\rm{A}^2_{\rm{model}}+\sigma_i^2/\rm{t}_i^\Gamma\right)^2},\nonumber\\
\rm{\hat{n}}&=& \sum_i e^{-i \frac{\rm{t}_0}{\Delta \rm{t}}} = \frac{1-\left(e^{-\rm{t}_0/\Delta \rm{t}}\right)^{\rm{n}}}{ 1-e^{-\rm{t}_0/\Delta \rm{t}}},\label{eq:awt}
\end{eqnarray}
where $\Delta$t is the time range between the very first and very last measurement of an individual quasar.

While a structure function with nonzero $\Gamma$ is ideal for characterizing quasars, it is simple enough to set $\Gamma = 0$ to parameterize the variability of fast periodically varying objects, some of which (like RR-Lyrae) could be confused with quasars. For these objects, V(t) is a constant V, and we can substitute $\Gamma = 0$ into Eq. \ref{eq:a2} to obtain 
\begin{equation}
\rm{V}^2=\rm{m}_0^2-\sigma_0^2+\frac{2}{\rm{V}^2_{\rm{model}}+\sigma_0^2}\left(\sigma_0^2 \rm{m}_0^2-\frac{1}{\rm{n}}\sum_i \rm{m}_i^2 \sigma^2_i\right)\label{eq:v2},
\end{equation}
and the weight
\begin{eqnarray}
\rm{weight}_{\rm{V}}=\frac{\rm{\hat{n}}}{2\rm{n}} \sum_i \frac{1}{\left(\rm{V}^2_{\rm{model}}+\sigma_i^2\right)^2}.\label{eq:vwt}
\end{eqnarray}

For either the quasar case or the fast variable case, our A$^2$ estimator and weight are only valid for one object and one small time bin. To use all the time information, we bin all time differences into 30 logarithmic bins covering the range from $10^{-2.0} = 0.01$ years to $10^{1.1} = 12.6$ years. The A$^2$ estimator for an individual source is then
\begin{eqnarray}
\rm{A}^2 &=& \frac{\sum_i \rm{A}^2(\rm{t}_i) \rm{weight}(\rm{t}_i)}{\sum_i \rm{weight}(\rm{t}_i)},\nonumber\\
\sigma^2_{\rm{A}^2} &=& \frac{1}{\sum_i \rm{weight}(\rm{t}_i)}\label{eq:a2ind}
\end{eqnarray}
where t$_i$'s are the different time bins. The $\chi^2$ goodness of fit is
\begin{eqnarray}
\chi^2 &=& \frac{\rm{A}^2_{\rm{model}}}{\rm{A}^2}\sum_i \left(\rm{A}^2(\rm{t}_i)- A^2\right) \rm{weight}(\rm{t}_i),\nonumber\\
\rm{N}_{\rm{DoF}} &=& \sum_i (\rm{weight}(\rm{t}_i) > 0).
\end{eqnarray}
The number of degrees of freedom, $\rm{N}_{\rm{DoF}}$, is not strictly correct, because our N(N-1)/2 time lags are taken from only N independent measurements. In addition, our weights in Eq. \ref{eq:awt} and Eq. \ref{eq:vwt} assume a typical quasar amplitude variation. For sources that do not vary, we significantly underestimate the weight and produce tiny $\chi^2$. We can roughly correct for this by multiplying $\chi^2$ by $\rm{A}^2_{\rm{model}}/\rm{A}^2$ as we do here. In general, we do not rely on this $\chi^2$ going to $\rm{N}_{\rm{DoF}}$, as a proper $\chi^2$ would. 

When studying populations of quasars, it is easy enough to calculate the ensemble A$^2$ for a time bin, t$_i$, as
\begin{equation}
 \rm{A}^2(\rm{t}_i) = \frac{\sum_j \rm{A}_j^2(\rm{t}_i) \rm{weight}_j(\rm{t}_i)}{\sum_j \rm{weight}_j(\rm{t}_i)}
\end{equation}
where j indexes our list of quasars. To calculate the uncertainty of a single bin in our ensemble average, we must account for the fact that each quasar has its own individual structure function, so the variance of A$^2$ is much larger than would be suggested by the error bars in Eq. \ref{eq:a2ind}. The variance divided by the number of quasars produces much more reasonable ensemble error bars than the inverse sum of weights. For some time bins at high time lag, there may be very few measurements, and the observed population variance can be zero or negligibly small. In these cases, we use the variance of all t$_i$ > 1 year measurements of A$^2$. Our ensemble error bars for individual time bins are then 
\begin{eqnarray}
\sigma^2_{\rm{A}_i^2}   &=& \max\left(\frac{\rm{Var}_i}{\rm{n}_i}, \frac{\rm{Var}_{\rm{t}_i > 1}}{\rm{n}_i}\right)\nonumber\\
\rm{Var}_i              &=& \frac{1}{\rm{n}_i-1}\left(            \sum_j \left(\rm{A}_{ij}^2\right)^2                - \frac{1}{\rm{n}_i}\left(\sum_j \rm{A}_{ij}^2\right)^2\right),\nonumber\\ 
\rm{Var}_{\rm{t}_i > 1} &=& \frac{1}{\rm{n}_{\rm{t}_i > 1}-1}\left(\sum_{\rm{t}_i > 1,j} (\rm{A}_{ij}^2)^2 - \frac{1}{\rm{n}_{\rm{t}_i > 1} }\left(\sum_{\rm{t}_i > 1, j} \rm{A}_{ij}^2\right)^2 \right).
\end{eqnarray}
Here, n$_i$ is the number of quasars that had at least one measurement pair in time bin t$_i$, and n$_{\rm{t}_i > 1}$ is the sum of all n$_i$'s for t greater than one year. 

Finally, to produce an ensemble structure function, we multiply each A$_i^2$ and $\sigma_{\rm{A}^2\ i}$ by t$_i^\Gamma$ (using a model $\Gamma$). We refit a single A$^2$ and $\Gamma$ to the resulting curve using a $\chi^2$ minimization routine in $\log$ A-$\log$ t space. We then iterate the process using the fit A$^2$ and $\Gamma$ as the model A$^2$ and $\Gamma$ until the two match. This also optimizes the fit. We then switch to the conventional V(t) = A t$^\gamma$, $\gamma = \Gamma/2$ for plotting and further analysis. We repeat this process across each of the \griz filters, which are taken to be independent.

\section{The Damped-Random Walk Model}\label{sect:drw}

The power law model we use to parameterize the ensemble structure function is purely empirical and not physically motivated. {Kelly} {et~al.} (2009); {Koz{\l}owski} {et~al.} (2010); {MacLeod} {et~al.} (2010) and others have fit the variation of individual quasars as a damped random walk (DRW) with a structure function parameterized as
\begin{equation}
\rm{V}(\rm{t}) = \rm{V}_\infty \left(1-e^{-t/\tau}\right)^{0.5}.\label{eq:drw}
\end{equation}

{Kelly} {et~al.} (2009) justifies Eq. \ref{eq:drw} as a physically reasonable model. The key features are that it is a classical random walk (setting the exponent to 0.5) with an exponential damping term with time scale $\tau$ that prevents the quasar from randomly walking by arbitrarily large values. {MacLeod} {et~al.} (2010) finds that $\tau$ is typically between 0.1 and 3 years and V$_\infty$ is typically between 0.1 and 0.5 mags. We discuss the DRW here only to provide context for why we use a power law structure function model. {MacLeod} {et~al.} (2012) and {Zu} {et~al.} (2012) offer more thorough statistical discussions of the DRW structure function, while {Mushotzky} {et~al.} (2011) and {Lovegrove} {et~al.} (2011) offer detailed time series analysis of very small quasar samples. 
%We study an analytic approximation of an ensemble DRW model to determine whether it is a significantly better fit to our data than a power law. 

The Damped Random Walk model of quasar variability has gained popularity in recent years, and it may be a superior model of quasar variability for well-sampled light curves. That being said, we employ the power law model for a variety of reasons. First, for light curves with only a few data points, the difference between the two fits is negligible. Second, ensemble structure functions are generally found to be consistent with power laws even if individual quasars have DRW variability. It is difficult to relate an individual  V$_\infty$ and $\tau$ to this power law, but assuming a constant $\gamma$ reduces individual quasar variability to a measurement of amplitude, A. This simplifying assumption facilitates the analysis in Section \ref{sect:varzl}. Finally, this work aims to parameterize variability for every source in the PS1-SDSS overlap. This is computationally reasonable for a simple power law fit, but would be prohibitively time consuming for a DRW fit, which generally require a Monte Carlo analysis for each light curve.

\section{Quasar Variability in the Observer Frame}\label{sect:varobs}

\begin{figure}[ht]
\includegraphics[width=0.49\columnwidth]{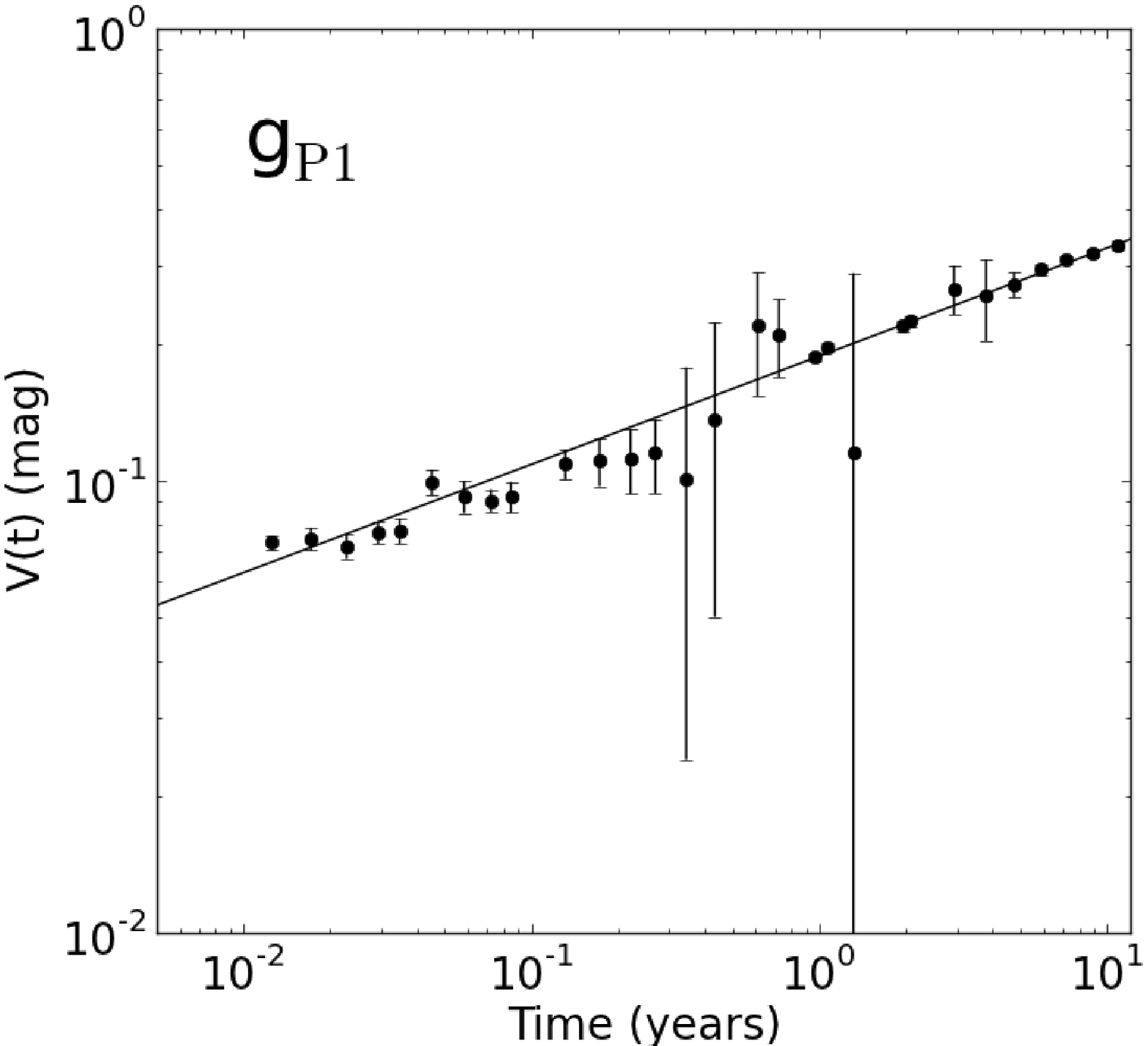}
\includegraphics[width=0.49\columnwidth]{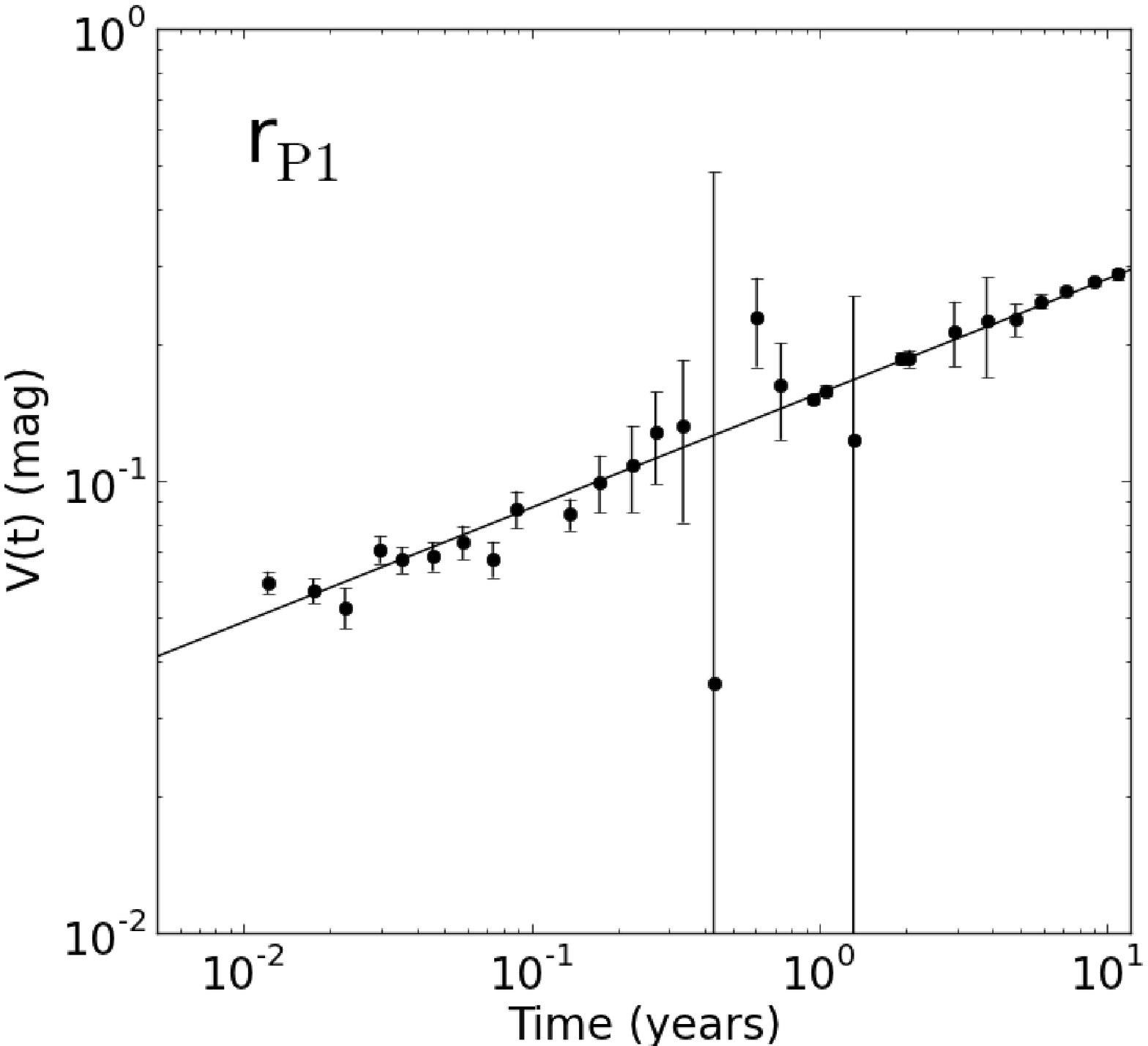}
\includegraphics[width=0.49\columnwidth]{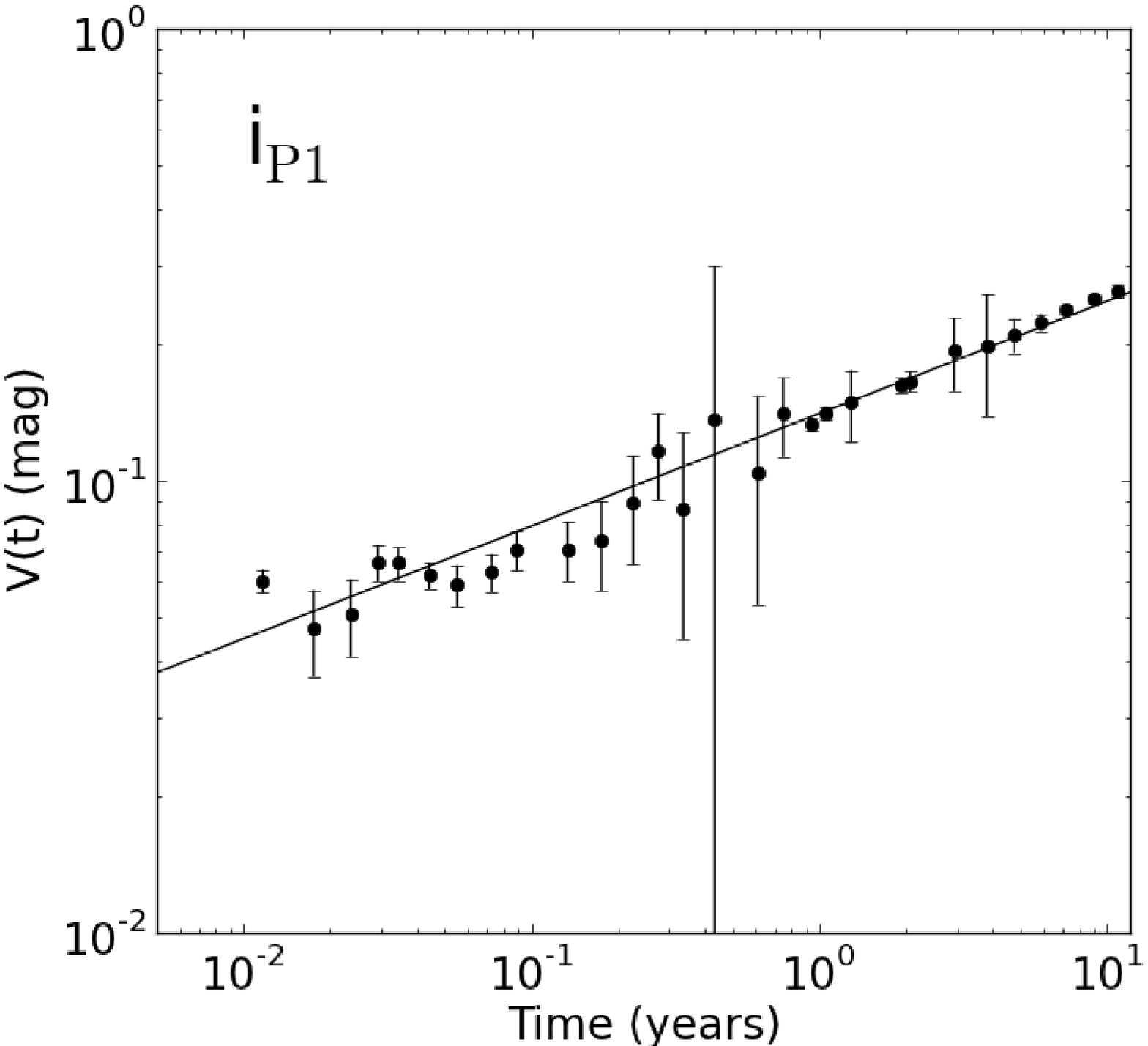}
\includegraphics[width=0.49\columnwidth]{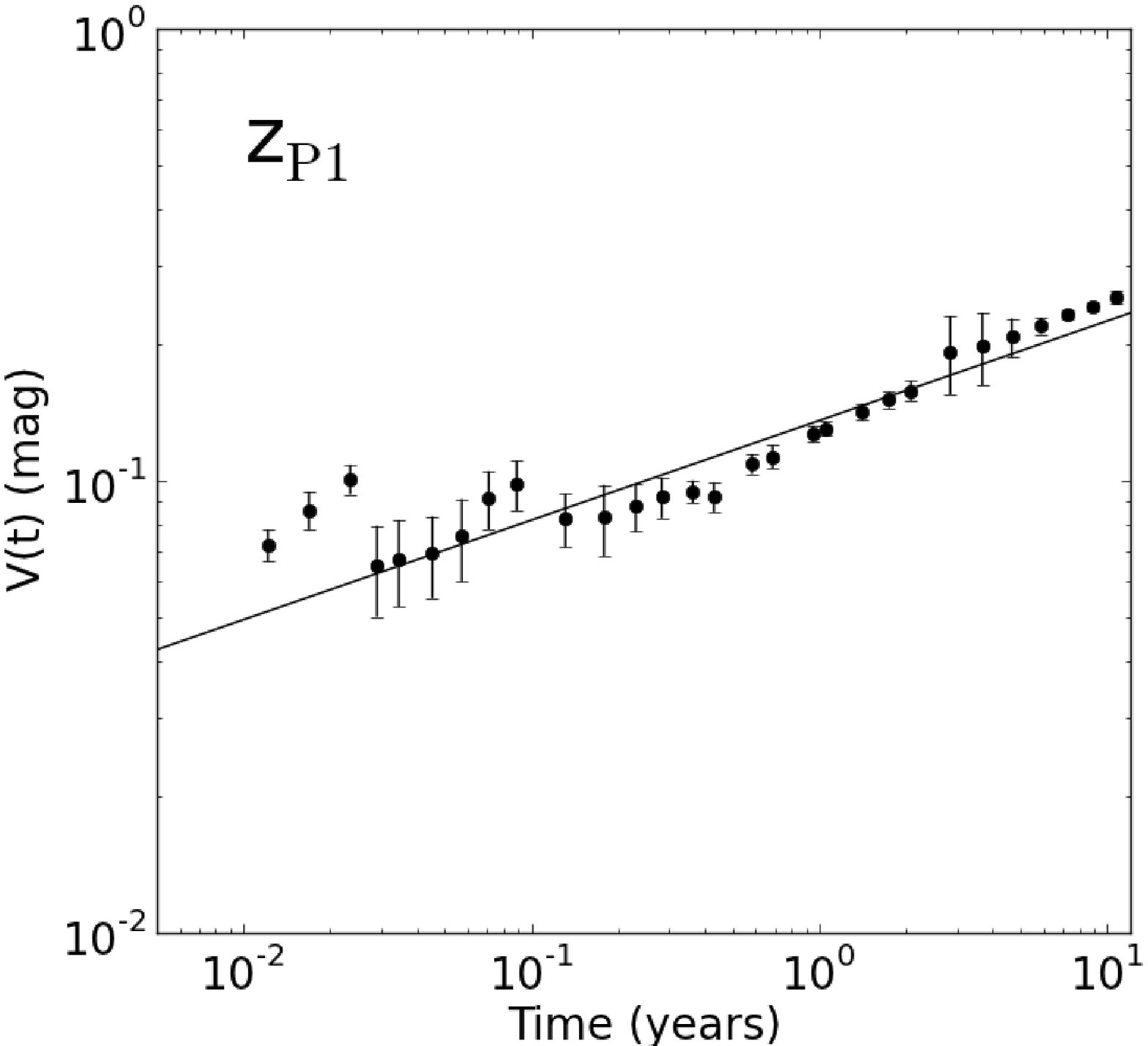}
\caption{\rm{The observer-frame ensemble structure function, V(t), of all quasars identified by {Shen} {et~al.} (2011) in \gps (upper left), \rps (upper right), \ips (lower left) and \zps (lower right).} } 
\label{fig:f3}\end{figure}

With the algorithm described in Section \ref{sect:var2}, we can perform the first of our science tasks: measuring the observer-frame variability of $10^5$ quasars. Ultimately, we are going to use our variability measurement of known quasars as a template for finding new quasars. So we must be able to distinguish a quasar from other sources by its variability without knowing its redshift or intrinsic luminosity. This includes using observer-frame times rather than correcting times by (1+z). We show our ensemble results in Table \ref{tab:vobsall} and Fig. \ref{fig:f3}. Many bins in Fig. \ref{fig:f3} have statistical large error bars, because they represent time lags of roughly 6 months, 18 months or 30 months, lags for which it is difficult to observe the same source. Reassuringly, we reproduce the general trend of quasar variability being largest for the bluer bands.  

\begin{table*}
\begin{tabular}{cccccccc}
        \hline
Filter &  N$_{\rm{quasars}}$ & Mean(N$_{\rm{obs}}$) & Offset & A                  & $\gamma$            & cov$_{\rm{A}\gamma}$ & $\chi^2_{\rm{red}}$  \\
        \hline
\gps   & 87725               & 6.93  & -0.0125  & 0.1899 $\pm$ 0.0014 & 0.2395 $\pm$ 0.0037 & -1.4e-06 & 1.13 \\ 
\rps   & 87401               & 7.05  & 0.0142  & 0.1574 $\pm$ 0.0015 & 0.2531 $\pm$ 0.0047 & -2.1e-06 & 0.87 \\
\ips   & 87860               & 7.03  & 0.0088  & 0.1430 $\pm$ 0.0016 & 0.2489 $\pm$ 0.0057 & -3.5e-06 & 1.26 \\ 
\zps   & 89917               & 7.31  & 0.0308  & 0.1426 $\pm$ 0.0016 & 0.2202 $\pm$ 0.0061 & -4.3e-06 & 4.45 \\
	\hline
\end{tabular}
\caption{\rm{The observer-frame variability parameters for the complete set of {Shen} {et~al.} (2011) quasars fit with a power law model. The "Offset" column is the average PS1-SDSS value for each filter. } }\label{tab:vobsall}
\end{table*}

We made several restrictions on these quasars to avoid biasing our PS1-SDSS comparisons. Some quasars have some extended source flux from the galaxy surrounding the quasars' black holes. The different PSFs from PS1 and SDSS would likely lead to a biased luminosity difference, so we only use sources that SDSS categorizes as point sources (SDSS type = `star'). PS1 does not have a uniformly applied star-galaxy separation statistic, so we require that the PS1 aperture magnitude be close to the PSF magnitude, i.e. mag$_{\rm{psf}}$ - mag$_{\rm{ap}} < 0.3$ in all filters. We also require that SDSS detect the object in the filter being measured as well as the $g'$ and $i'$ filters required for our PS1-SDSS conversion in Eq. \ref{eq:spcon}. The $g'$ band requirement rejects many $z > 3$ quasars. When we study the redshift dependence of quasar variability, we do not extend any analysis into that redshift range. Finally, we require that there be at least 1 PS1 measurement of the quasar with the filter in question. The number of quasars being studied in each filter is listed as N$_{\rm{quasars}}$ in Table \ref{tab:vobsall}. The average number of PS1 measurements in each filter of quasars being studied in that filter is listed as Mean(N$_{\rm{obs}}$). A typical quasar has 7 observations per filter from PS1. This small number of data points is another reason we do not try to fit $\Gamma$ for each quasar in each filter. Combining PS1 with SDSS is useful to get more time information, and our variability characterizations of individual quasars tend to be weak. 

Table \ref{tab:vobsall} and Fig. \ref{fig:f3} have some important qualitative implications for selecting quasars by their variability. The amplitude of variability at 1 year, A, is 0.1899 magnitudes in \gps. This makes it very difficult to use variability to select quasars in surveys or redshift ranges where statistical or calibration uncertainty is $\geq 0.1$. Additionally, using the structure function in Eq. \ref{eq:v}, we can project our fit down to the three day scale where typical variability is 0.060 magnitudes or to the 10 year scale where it is 0.330 magnitudes. Using a small number of measurements over a decadal time scales can thus be more effective than using many measurements over a small time period. 

\section{The Average Magnitude Offset Between PS1 and SDSS Measurements of Quasars}\label{sect:offset}

When comparing PS1 and (converted) SDSS magnitudes of known quasars, we noticed an unexpected phenomenon: quasars, on average, apparently became dimmer by the amount listed in the "Offset" column in Table \ref{tab:vobsall}. This effect is strongest in the bluer filters, making it superficially consistent with physical variation.

We considered five possible origins for this effect: (1) a few outliers skewed the mean, (2) PS1-SDSS differences in measuring (barely) extended sources, (3) an astrophysical tendency of quasars to get dimmer on decadal time scales, (4) bias from the initial SDSS quasar selection and (5) our conversion in Eq. \ref{eq:spcon} producing biased offset for quasars. To check against the "outlier" hypothesis (1), we used the mean, median and outlier rejected mean PS1-SDSS offset. To check against the "extended source" hypothesis (2), we used quasars with mag$_{\rm{psf}}$ - mag$_{\rm{ap}}$ < 0.1, a stricter "point source" requirement. The effect was essentially unchanged in both cases.

To check against the physical variation hypothesis (3), we examined the average magnitude difference versus time using only PS1 data. There was no measurable trend. In addition, the PS1-SDSS difference is independent of time lag, although we can only measure it with statistical significance for time lags greater than two years.

\begin{table}
\begin{tabular}{cccc}
        \hline
Filter &  Offset$_{\rm{SDSS}}$ & Offset$_{\rm{PS1}}$ & Offset$_{\rm{Both}}$  \\
        \hline
\gps   & 0.0425                & -0.0628             & -0.0010\\
\rps   & 0.0163                & -0.0739             & -0.0224\\
\ips   & 0.0275                & -0.0528             & -0.0084\\
\zps   & 0.0234                & -0.0498             & -0.0125\\
        \hline
\end{tabular}
\caption{\rm{We show the average PS1-SDSS difference for an SDSS 17 < $g'$ < 18 quasar sample (Offset$_{\rm{SDSS}}$) and a PS1 17 < \gps < 18 quasar sample (Offset$_{\rm{PS1}}$). The Residual column is just the average of the two offset and estimates the residual bias after accounting for selection bias. The SDSS magnitudes here are SDSS magnitudes converted into the PS1 system.} }\label{tab:offsets}
\end{table}

To check against SDSS selection bias (4), we measured the average PS1-SDSS offsets across four bands in three quasar subpopulations. The first population is those quasars with SDSS magnitudes 17 < $g'$ <18. The second is those quasars with PS1 magnitudes 17 < \gps < 18. The third population satisfies both 17 < $g'$ <18 and 17 < \gps < 18. The lower bound of 17 was chosen to avoid any saturation effects in both surveys at roughly g = 15. The upper bound of 18 was chosen to be significantly brighter than the SDSS quasar spectroscopy selection limiting magnitude of roughly 20.2. Only a miniscule fraction of quasars will have varied by more than 2 magnitudes, so our spectroscopic sample should be complete in this magnitude range. In Table \ref{tab:offsets}, we see that populations 1 and 2 have opposite (though not equal) offsets. Setting a limiting magnitude clearly biases our measurements in the bright direction. Since the third population has symmetric SDSS and PS1 constraints, we would expect the offset to be nearly zero, and indeed the offset of a symmetrically chosen sample is on average smaller than that of the either two samples. The residual offset indicates that there is a second reason that average SDSS-PS1 differences are significantly nonzero, likely filter effects (5).

\begin{figure}[ht]
\includegraphics[width=0.49\columnwidth]{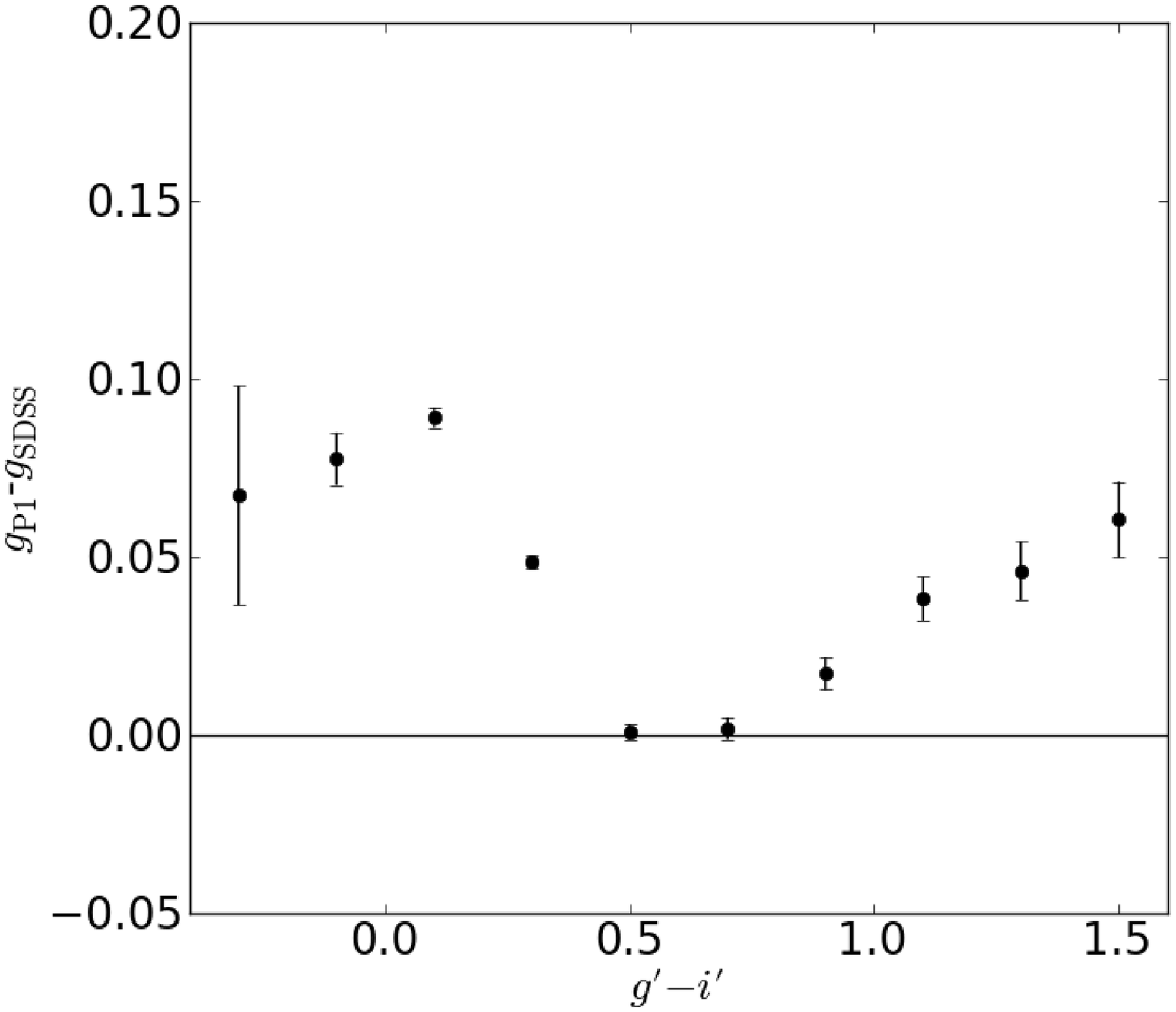}
\includegraphics[width=0.49\columnwidth]{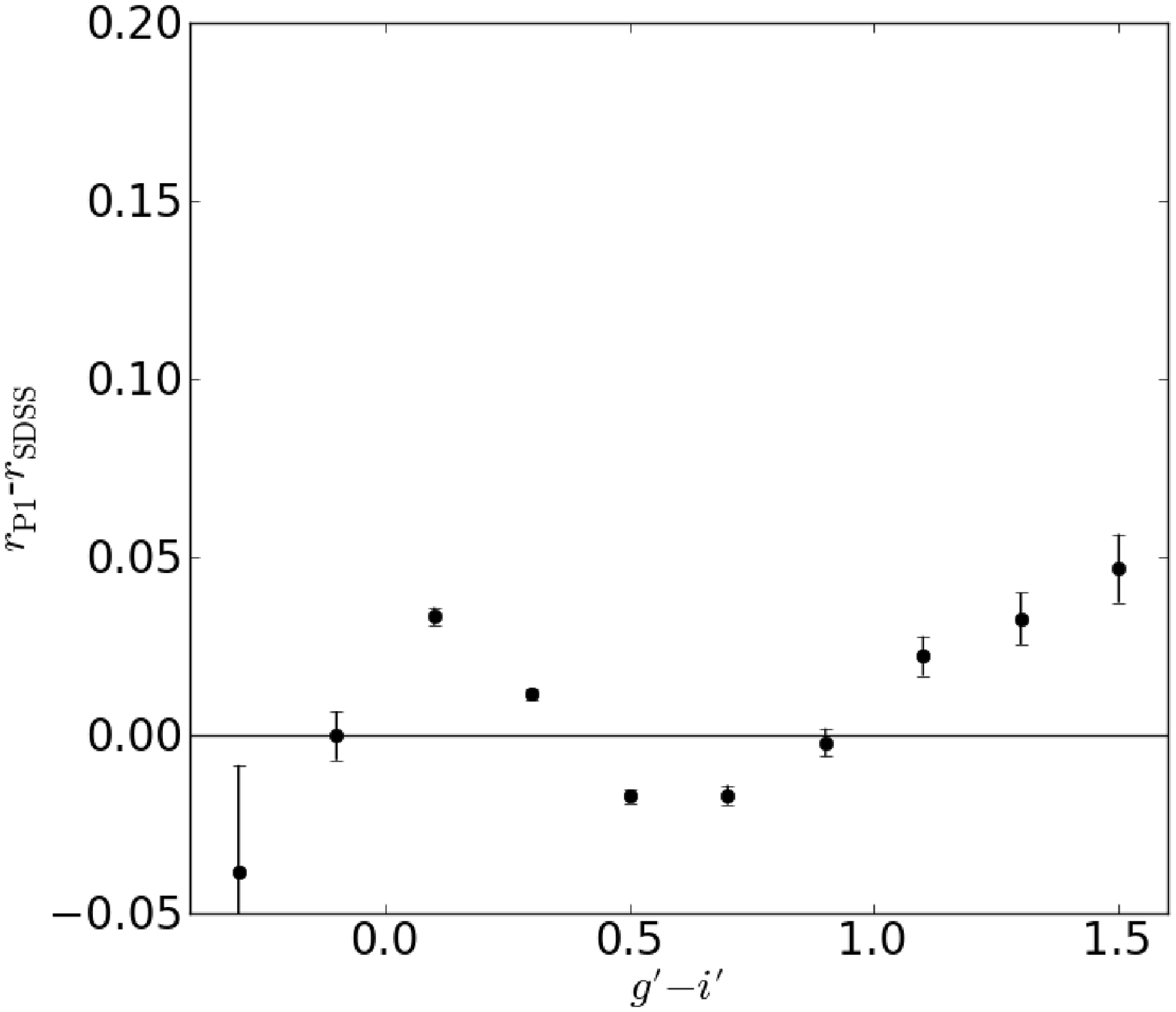}
\includegraphics[width=0.49\columnwidth]{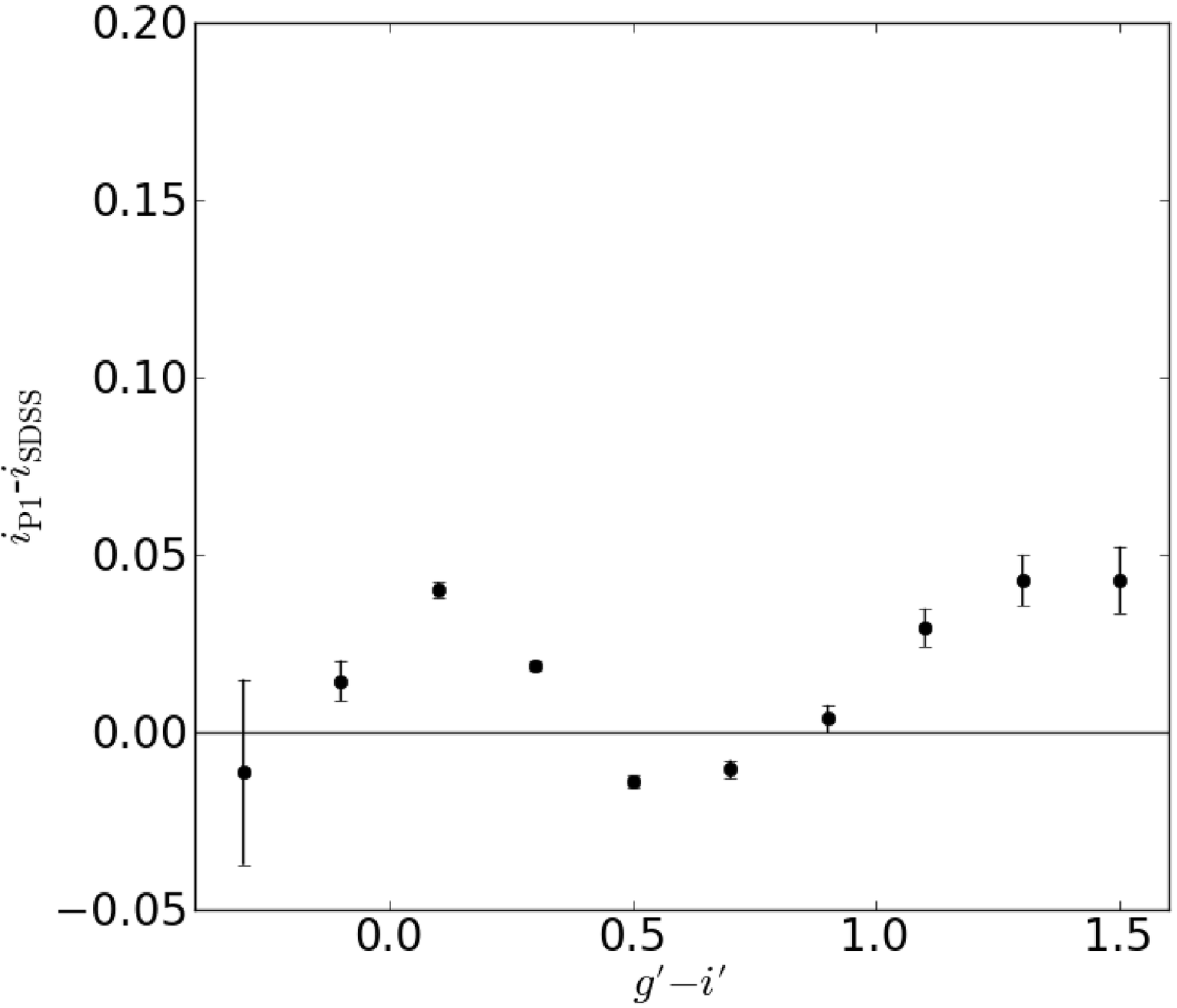}
\includegraphics[width=0.49\columnwidth]{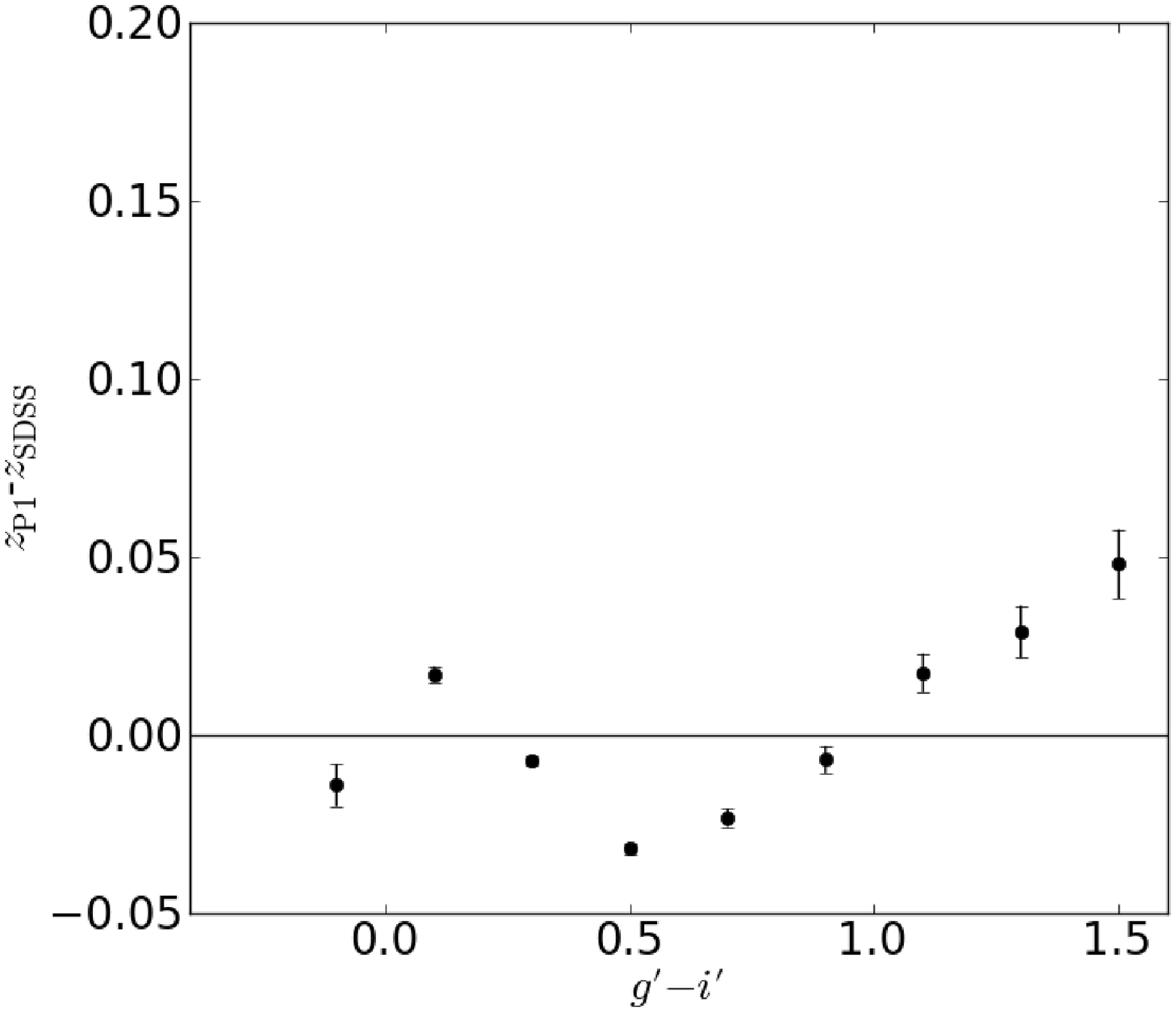}
\caption{\rm{The average PS1-SDSS offset (using converted SDSS magnitudes) versus $g' - i'$ in \gps (upper left), \rps (upper right), \ips (lower left) and \zps (lower right).} }
\label{fig:f4}\end{figure}

We investigated how the Offset term varies versus $g' - i'$ in Fig. \ref{fig:f4}. If the conversion from Eq. \ref{eq:spcon} worked well for quasars, the Offset term would be constant. But instead there is clear color dependency on the offset term. The Offset terms tend to be particularly large around $g' - i' = 0.2$, the median quasar value. We conclude that the conversion in Eq. \ref{eq:spcon} does not work perfectly well for quasars and that this is the likely cause of our residual offset. This is not surprising since Eq. \ref{eq:spcon} was derived for confirmed stars and not quasars. A separate filter transformation could be derived for quasars, but since observer-frame quasar spectra vary greatly it would likely have large scatter. More importantly, this quasar filter transform would not work well for stars and would produce apparent variability in static objects.

When studying the rest-frame variability of known quasars in Section \ref{sect:varrest}, we add the "Offset" term from Table \ref{tab:vobsall} to our SDSS magnitude to produce a more precise measurement of quasar variability. However, we do not use the "Offset" term when measuring observer-frame variability. This would add a great deal of false variability to otherwise static sources for which the conversion in Eq. \ref{eq:spcon} works. Ultimately, the variability of quasars on the multi-year time scales that separate SDSS and PS1 measurements is much larger than this offset, so we are still able to produce sensible variability results, despite this offset. Reassuringly, we do not see any "jump" in Fig. \ref{fig:f3} at the time = 2 years region where we transition from PS1-PS1 variability to PS1-SDSS variability.

\section{Variability Selection of Quasar Candidates}\label{sect:varsel}

Several groups ({Eyer} 2002; {Koz{\l}owski} {et~al.} 2010; {Schmidt} {et~al.} 2010; {Butler} \& {Bloom} 2011; {Kim} {et~al.} 2011; {MacLeod} {et~al.} 2011; {Palanque-Delabrouille} {et~al.} 2011) have produced criteria for selecting quasars using variability. But these groups all used relatively small surveys with many observations and focused on the single filter variability. In Table \ref{tab:vobsall}, we see that we typically have only 7 observations per object per filter. This means that we often cannot robustly fit anything beyond a variability amplitude in a single filter. The variability selection method we present here distinguishes quasars from other objects even if we only have a small number of observations. 

A simple approach to selecting quasars with variability is to reduce all the information about source variability down to a two variable schema for identifying objects that vary in a quasar-like way. In principle, we could cross-match measurements from different filters at different times to get more time information. But this would require us to implement a short term color variability model that could get significant information from non-simultaneous measurements from different filters. For simplicity, our method just combines the four single filter results into one weighted mean result. For each quasar candidate, we calculate two quantities: the 1 year amplitude of variability assuming a structure function model, A, and the difference between the quasar goodness of fit (with nonzero $\gamma$) and the RR-Lyrae goodness of fit (with $\gamma = 0$), $\Delta \chi^2 = \chi^2_{\rm{qso}}$-$\chi^2_{\rm{rrl}}$. A negative $\Delta \chi^2$ indicates that an object varies more on long time scales, while a positive $\Delta \chi^2$ is consistent with an object that varies in a fast, periodic way. Making cuts in this 2D space allows us to distinguish between quasars, non-varying objects and quickly varying object like RR-Lyrae. 

Our formulae for A$^2$ from Eq. \ref{eq:a2} and V$^2$ from Eq. \ref{eq:v2} apply to single filter measurements. To produce a single A$^2$ and V$^2$, we average the A$^2$ and V$^2$ from each band, weighted by the average values of A$^2$ for each filter list in Table \ref{tab:vobsall}. We index these average values as A$^2_f$ where $f = \grizy$. The mean A$^2$ and V$^2$ for a quasar candidate is then
\begin{eqnarray}
\rm{A}^2 &=& \sigma^2_{\rm{A}} \sum_f \frac{\rm{A}^2_f}{\rm{A}^2_{0f}  \sigma^2_{\rm{A}f}}\\
\sigma^2_{\rm{A}} &=& \left(\sum_f \frac{1}{\rm{A}^2_{0f} \sigma^2_{\rm{A}f}}\right)^{-1}\nonumber\\
\rm{V}^2 &=& \sigma^2_{\rm{V}} \sum_f \frac{\rm{V}^2_f}{\rm{A}^2_{0f} \sigma^2_{\rm{V}f}}\\
\sigma^2_{\rm{V}} &=& \left(\sum_f \frac{1}{\rm{A}^2_{0f} \sigma^2_{\rm{V}f}}\right)^{-1}\nonumber\\
f &=& \griz\label{eq:amean}.
\end{eqnarray}
From this point on, when discussing quasar selection, we take "A" and "V" to be this weighted mean A and V. Note that A$^2$ is normalized by the single band ensemble average values of A$^2_{0f}$. Quasars' average A$^2$, A$^2_{\rm{model}}$, is thus normalized to 1. The ability to average variability amplitudes across filters is a major advantage of using a power law structure function with a fixed $\gamma$. Combining four damped random walk structure functions or four structure functions with a variable $\gamma$ would not be more complicated. 

We can also define a combined, multiband goodness of fit, $\chi^2$ as
\begin{eqnarray}
\chi_{\rm{qso}}^2 &=& \frac{1}{\rm{A}^2}\sum_{f, i} \left(\rm{A}_f^2(\rm{t}_i)- A^2 \rm{A}^2_{0f}\right) \rm{weight}_f(\rm{t}_i),\nonumber\\
\chi_{\rm{rrl}}^2 &=& \frac{1}{\rm{A}^2}\sum_{f, i} \left(\rm{V}_f^2(\rm{t}_i)- V^2 \rm{A}^2_{0f}\right) \rm{weight}_f(\rm{t}_i),\nonumber\\
\Delta \chi^2 &=& \chi^2_{\rm{qso}}-\chi^2_{\rm{rrl}}
\end{eqnarray}

Here we are summing over all four bands with $f$ and all thirty time bins, denoted by t$_i$. A$^2$ and V$^2$ have slightly different weights defined in Eq. \ref{eq:awt} and Eq. \ref{eq:vwt}, respectively. Note that we multiplied our $\chi^2$'s by the analogous A$^2_{\rm{model}}$/A$^2$ factor from Eq. \ref{eq:awt} and Eq. \ref{eq:vwt}, but A$^2_{\rm{model}}$ = 1. Due to the relatively small number of observations we make of the average quasar, the $\chi^2$ itself has a large variance and is not useful as a selection variable, but the difference, $\Delta \chi^2$, is more robust.  

To examine the usefulness of A-$\Delta \chi^2$ space for selecting quasars, we define five test datasets. The first is just our set of quasars from {Shen} {et~al.} (2011). 

The second sample is a set of PS1-SDSS point sources (sources marked as type `star' by SDSS) with a broad quasar color cut:
\begin{eqnarray}
-0.2 &<& g'-r' < 0.9\nonumber\\
-0.2 &<& r'-i' < 0.6\nonumber\\
-0.15 &<& i'-z' < 0.5\nonumber\\
i' &<& 20.1\label{eq:colorbox}
\end{eqnarray}
which we take from {Schmidt} {et~al.} (2010). This is a reasonable quasar color selection for a region where there is no u band data. We limit our sources to $i'< 20.1$ so that they have a similar magnitude distribution as the {Shen} {et~al.} (2011) quasars, and we only use the 203,892 sources with RA$ < 6^{hr}$, $-2^\circ < $Dec$< 2^\circ$  to make the sample more manageable. 

Our third sample is the 483 RR Lyrae from {Sesar} {et~al.} (2010). RR Lyrae are an interesting test case because their variability is of similar amplitude to quasar variability, but is regular and occurs on time scales of days. 

It would be difficult to estimate the purity and completeness of a variability-selected quasar sample with a mix of spectroscopically selected quasars and photometrically selected candidates. If we restrict ourselves to quasar candidates that satisfy Eq. \ref{eq:colorbox} and are also detected in the Wide-field Infrared Survey Explorer (WISE, {Wright} {et~al.} 2010) survey, we can produce fairly pure and complete quasar samples. {Wu} {et~al.} (2012) shows that requiring $z'-W1 > 0.66(g'-z') +2.01, i' < 20.5$ produces a quasar sample that is roughly 98.3\% complete and 95.6\% pure. We call the sources which pass these cuts, our fourth and fifth populations, "SDSS-WISE Quasars" and those that fail "SDSS-WISE Stars". 

\begin{figure}[ht]
\includegraphics[width=0.99\columnwidth]{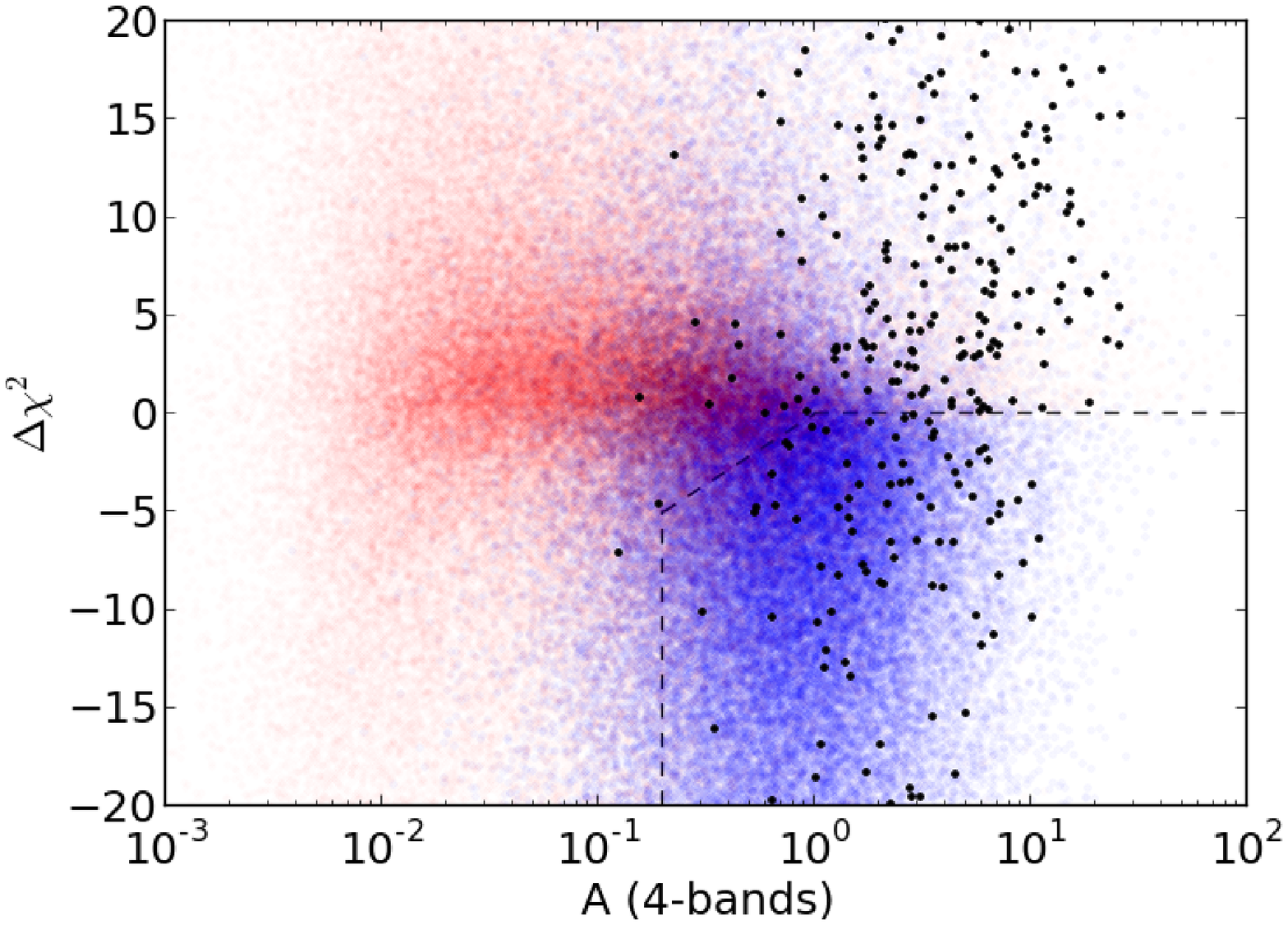}
\includegraphics[width=0.99\columnwidth]{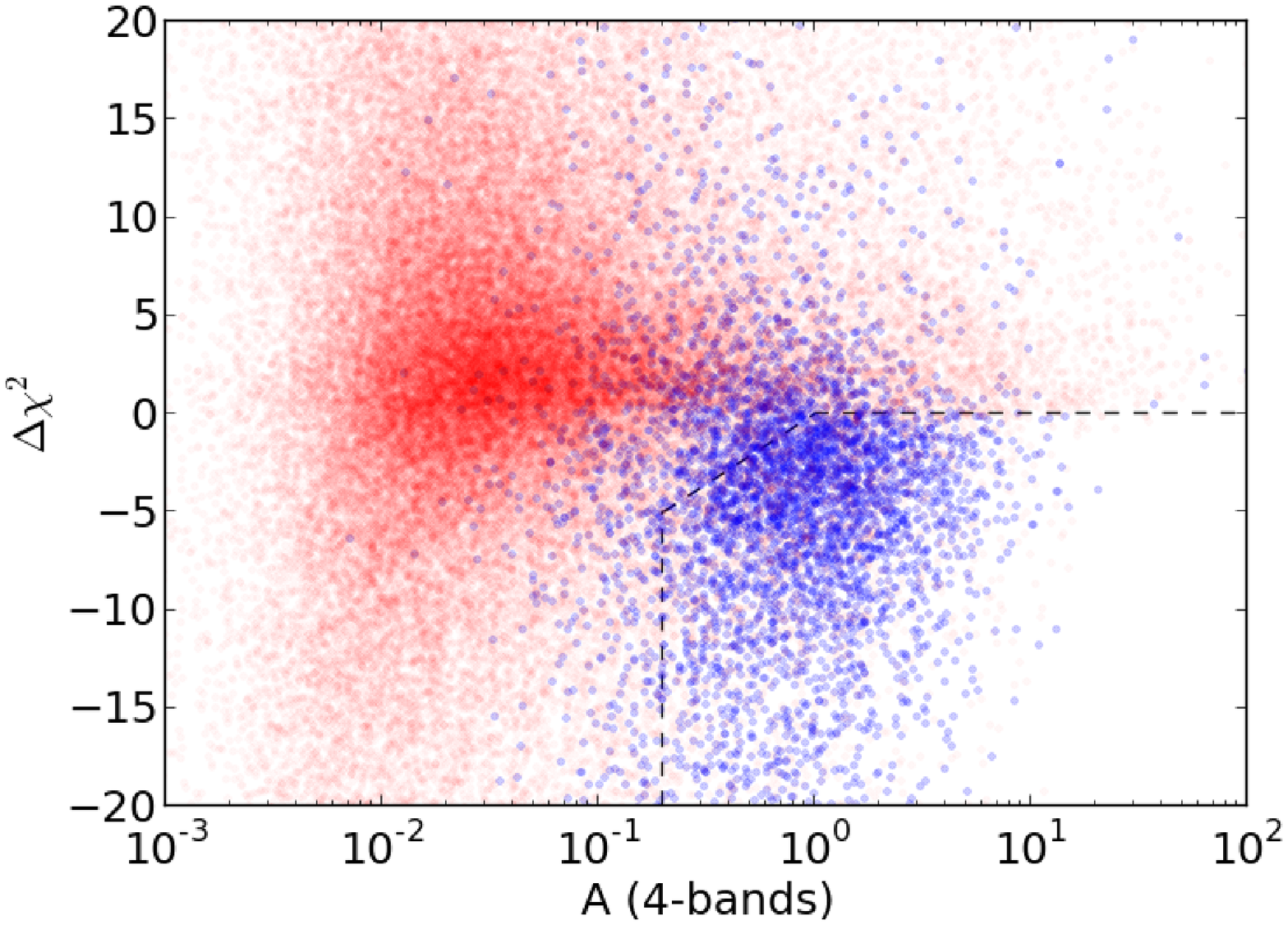}
\caption{\rm{Top: the A-$\Delta \chi^2$ distribution of spectroscopic quasars (blue), quasar-colored objects (red) and RR Lyrae (black). Bottom: SDSS-WISE matches that pass the $griz$ quasar cuts with SDSS-WISE quasars in green and SDSS-WISE stars in red. The transparency of each population is normalized for visual clarity.} } 
\label{fig:f5}\end{figure}

For all five samples, we only use sources that have a pair of measurements (in one band) with $0.01\ \rm{year} < \Delta t < 0.8\ \rm{year}$ and another pair with $\Delta t > 2\ \rm{year}$. This requirement ensures that we study the objects on both relatively short and relatively long time scales. Currently 85 \% of spectroscopically confirmed quasars and 77\% of RR Lyrae satisfy this requirement. We show the distribution of each of the five populations in A-$\Delta \chi^2$ space in Fig. \ref{fig:f5}.

To select for quasars in A-$\Delta \chi^2$ space, we adopt a set of cuts in A-$\Delta \chi^2$ space: 
\begin{equation}
\rm{A} > 0.2,\ \Delta \chi^2 < 0, \Delta \chi^2 < 7.2 \log_{10} \rm{A}  \label{eq:cuts}.
\end{equation}
In Fig. \ref{fig:f5} we see that the horizontal cut from Eq. \ref{eq:cuts} distinguishes quasars from RR Lyrae and the vertical and diagonal cuts distinguish quasars from quasar-colored stars. 

\begin{table}
\begin{tabular}{cccc}
        \hline
Population      &  Total Number & Number Pass & Fraction Pass\\
        \hline
Spectroscopic Quasars   & 85360  & 60274 & 0.706\\
PS1 Photometric Quasars & 203892 & 16433 & 0.081\\
Spectroscopic RR-Lyrae  & 383    & 123   & 0.321\\ 
SDSS-WISE Quasars       & 5202   & 3484  & 0.670\\
SDSS-WISE Stars         & 119409 & 3742  & 0.041\\
	\hline
\end{tabular}
\caption{\rm{A numerical description of in Fig. \ref{fig:f5}. For each of our five populations, we show the total number studied, the number which pass our quasar cut and the fraction that pass our quasar cut. } }\label{tab:cuts}
\end{table}

We see that even with a small amount of data, variability alone is a moderately effective method for selecting quasars when no u-band is available. Our proposed cut recovers 71\% of known quasars while rejecting 92\% of quasar-colored objects and 68\% of RR-Lyrae. Projecting our cut results over the entire SDSS area, we would select approximately 800,000 candidates. 60,274 of these would be known spectroscopically confirmed quasars, and we would expect several times this number would be new quasars. To estimate purity more precisely, we examine the SDSS-WISE candidates. Ignoring the slight incompleteness and impurity of these samples, we find that our selection list is 67\% complete and 48\% pure. The initial sample, before variability selection, was 4.1\% pure. 

Our variability selection method cannot compete with u-band or even SDSS-WISE quasar selection over the broad population of quasars. But variability will be an effective method of quasar selection across the roughly 15,000 square degrees of PS1 area with no SDSS coverage, can be used to find quasars past the WISE limit and will be effective even in redshift ranges where pure photometric selection fails. In addition, variability selection can be used to complement any photometric selection method. 

Other groups with more highly sampled data can produce pure samples with higher completeness. {Schmidt} {et~al.} (2010) produces samples that are estimated to be 96\% pure and 90\% complete in stripe 82 without using a $u'$ band by selecting in A-$\gamma$ space. But our dataset fundamentally lacks this level of information on a per-quasar basis. In many cases, we would not be able to fit a $\gamma$ with any precision. As we obtain more epochs in PS1, using different methods may become more viable. 

\section{Quasar Variability in the Rest-Frame}\label{sect:varrest}

\begin{figure}[ht]
\includegraphics[width=0.49\columnwidth]{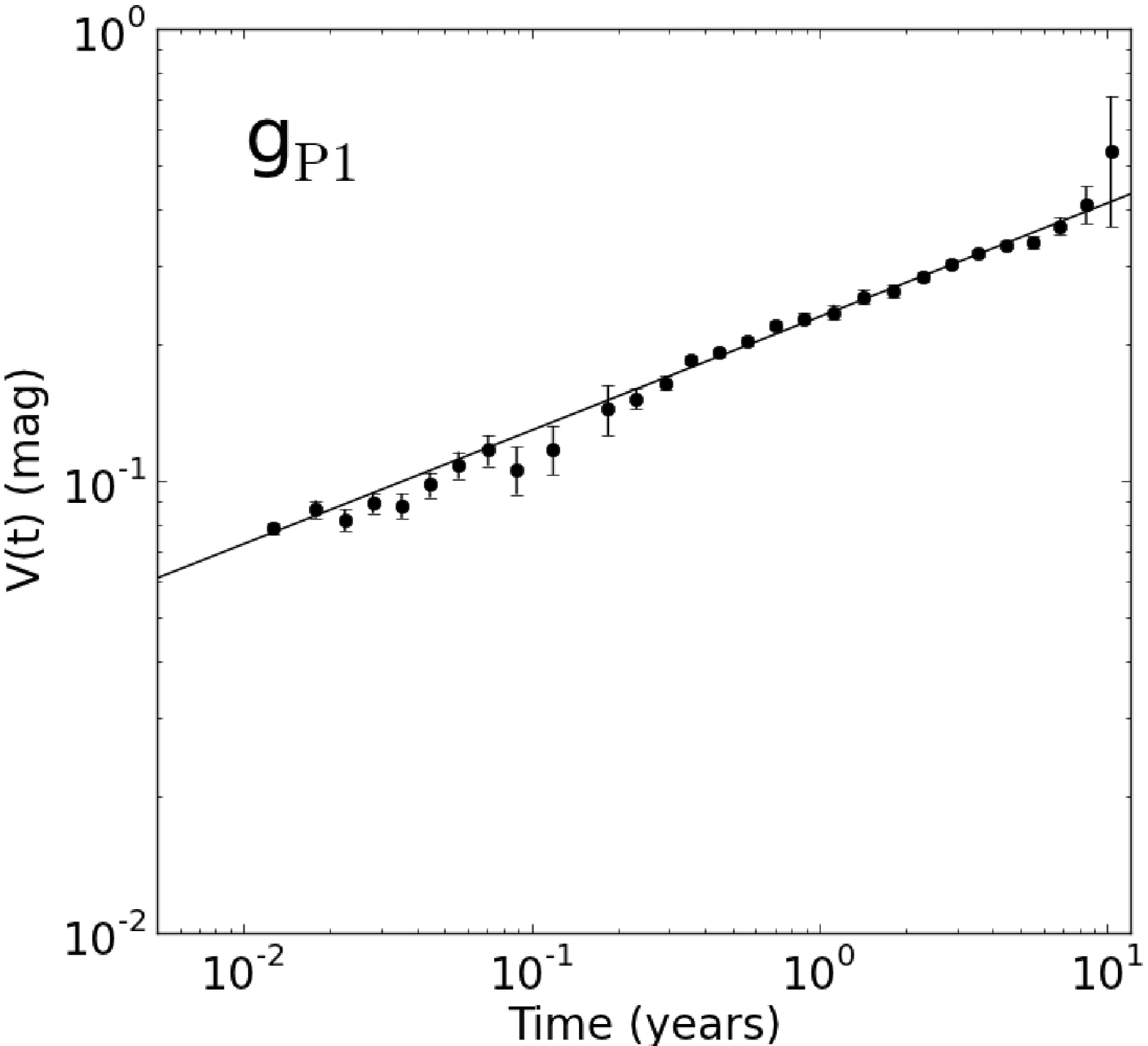}
\includegraphics[width=0.49\columnwidth]{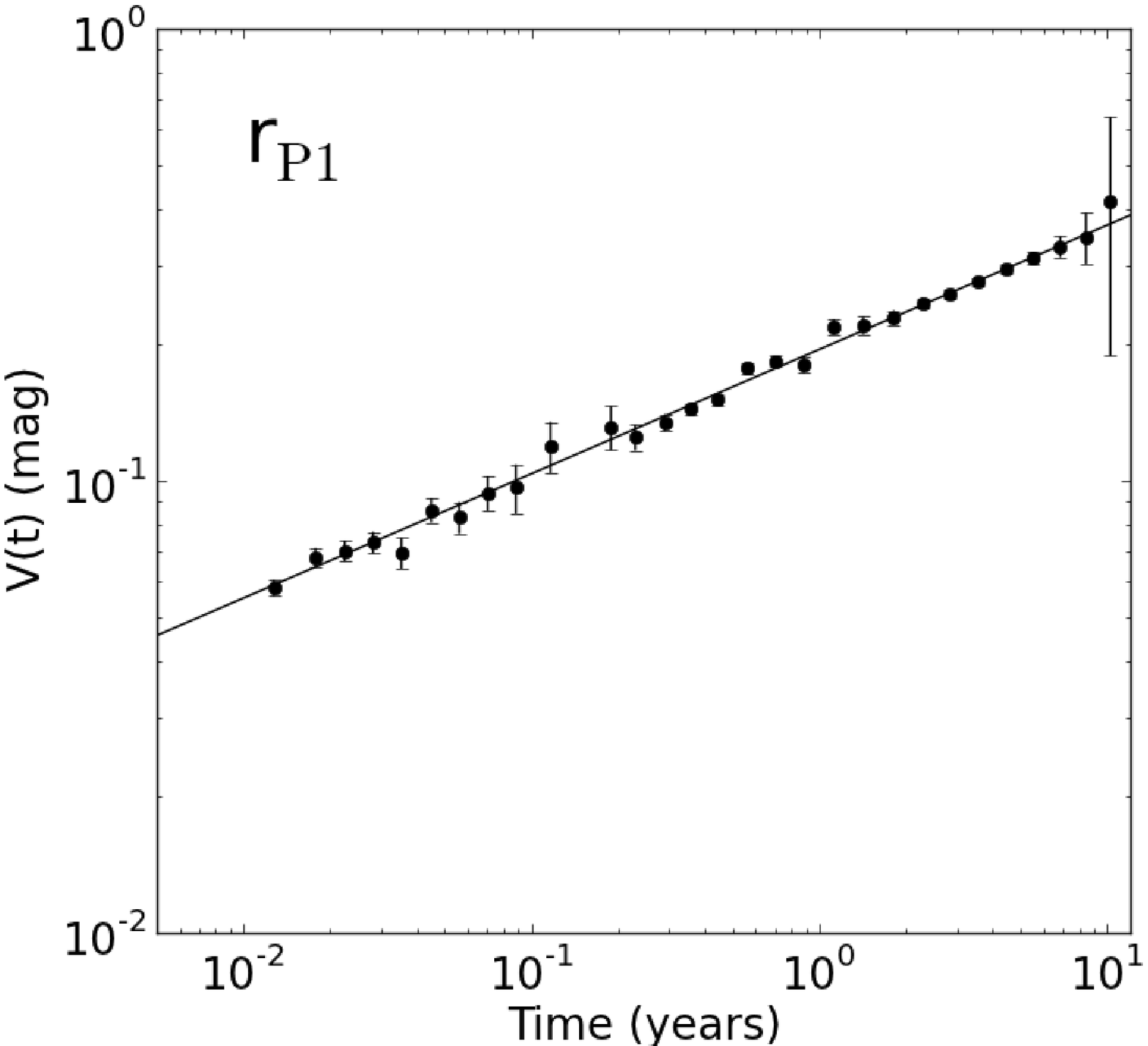}
\includegraphics[width=0.49\columnwidth]{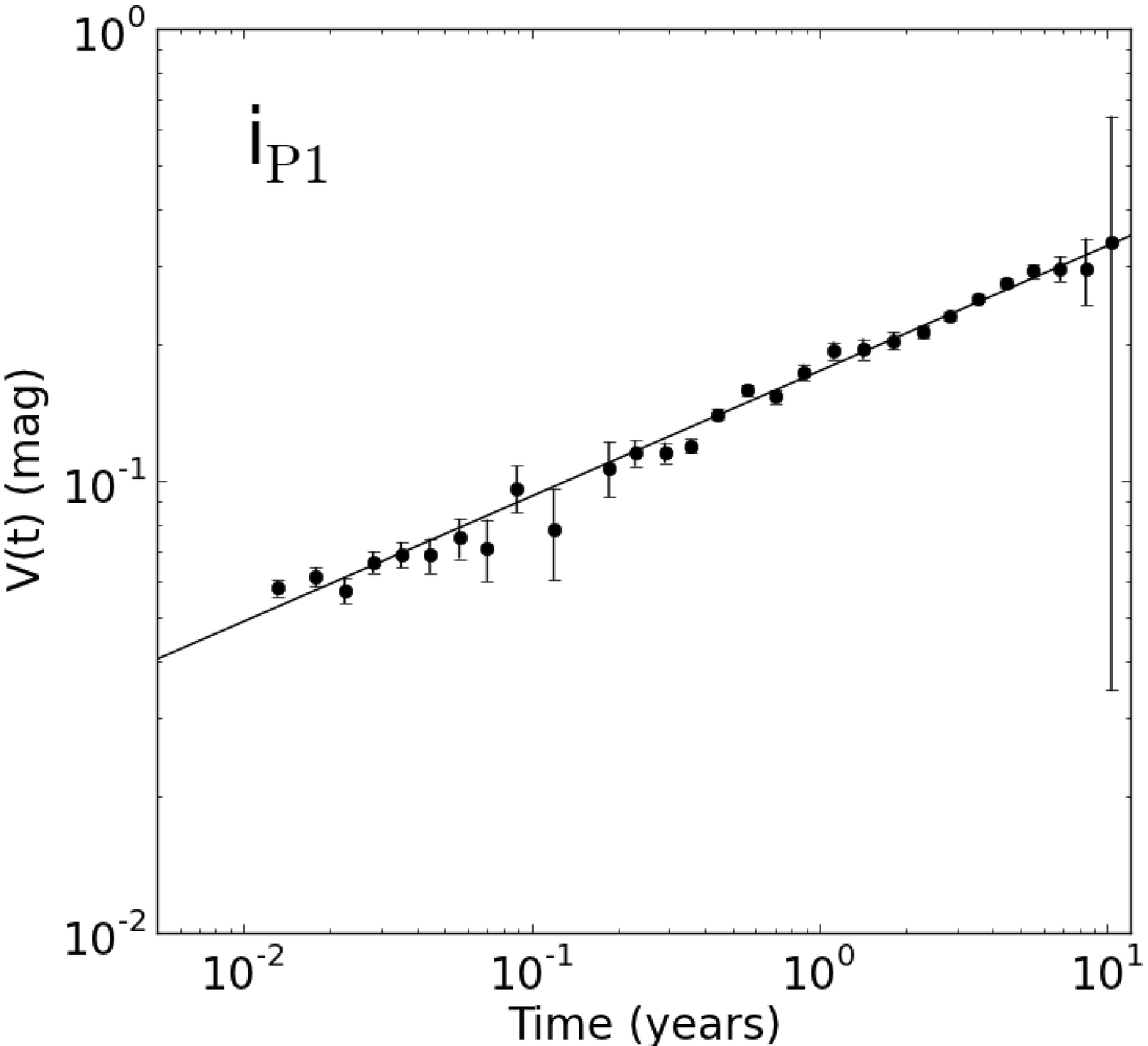}
\includegraphics[width=0.49\columnwidth]{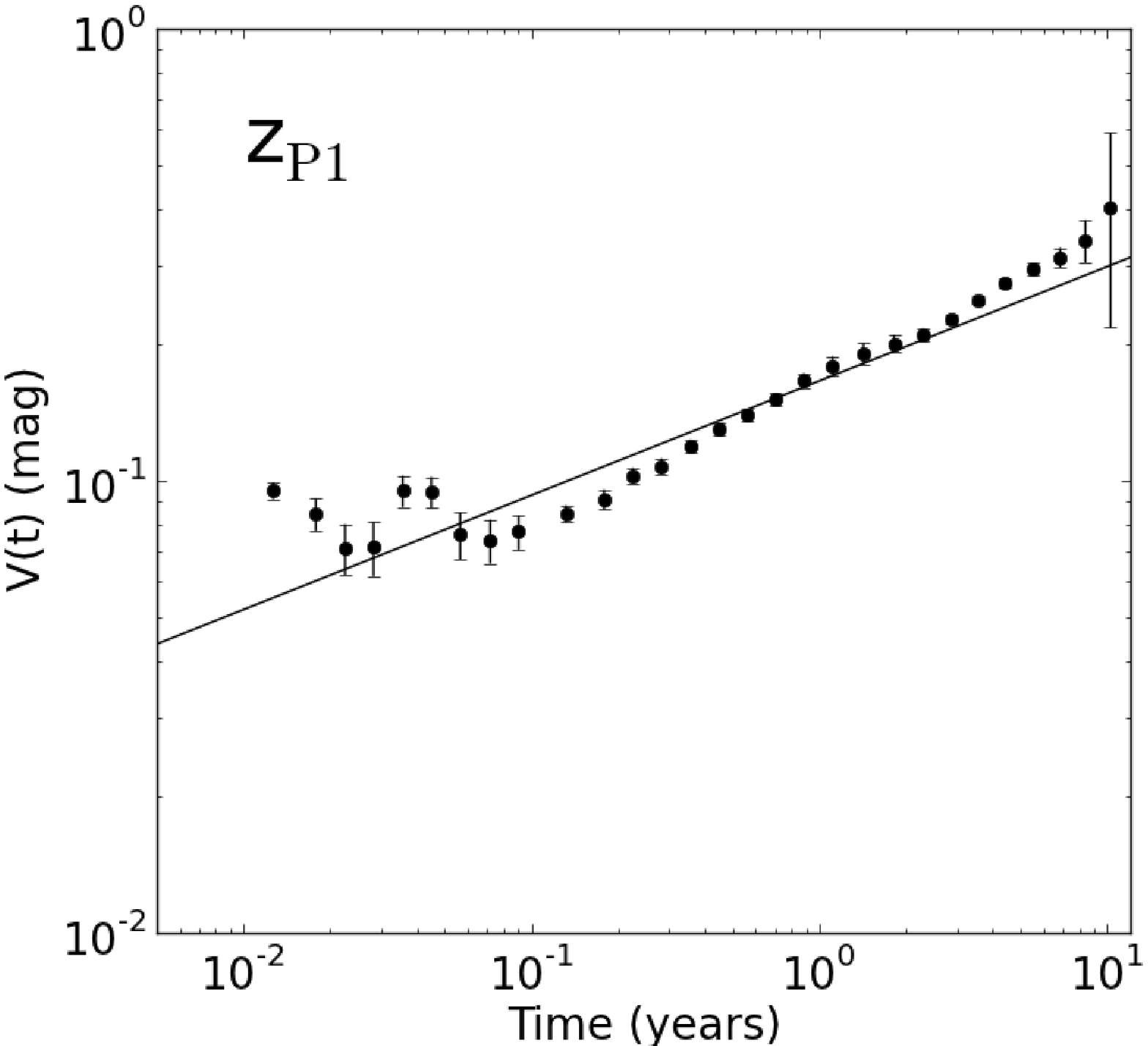}
\caption{\rm{The rest-frame ensemble structure function, V(t), of all quasars identified by {Shen} {et~al.} (2011) in \gps (upper left), \rps (upper right), \ips (lower left) and \zps (lower right). Here, we fit the data with a power law structure function.} } 
\label{fig:f6}\end{figure}

While using PS1-SDSS quasar variability to find new quasars has a promising future, the current PS1-SDSS quasar catalog also has the statistical power to measure rest-frame ensemble quasar variability with great precision. To do this accurately, we correct for time dilation and the PS1-SDSS "Offset" mentioned in Section \ref{sect:offset}. Since the quasars from {Shen} {et~al.} (2011) all have precise redshifts, we divide all measurements times by $(1+z)$ so that time lags are in the quasars' rest-frame. We also add the PS1-SDSS offset from Table \ref{tab:vrestall} to the SDSS magnitude and add an additional 0.01 magnitude of uncertainty to the SDSS magnitude to account for residual uncertainty after the "Offset" is added as discussed in Section \ref{sect:offset}. With these slight modifications, we measure rest-frame variability using the binning and fitting techniques from Section \ref{sect:var}. The results of these fits are in Table \ref{tab:vrestall} and Fig. \ref{fig:f6}.

\begin{table*}
\begin{tabular}{cccccccc}
        \hline
Filter &  N$_{\rm{quasars}}$ & Mean(N$_{\rm{obs}}$) & Offset & A                   & $\gamma$            & cov$_{\rm{A}\gamma}$ & $\chi^2_{\rm{red}}$  \\
        \hline
\gps   & 87725               & 6.93                 & -0.0125 & 0.2314 $\pm$ 0.0013 & 0.2514 $\pm$ 0.0036 & 2.4e-07 & 1.19 \\ 
\rps   & 87401               & 7.05                 & 0.0142 & 0.1959 $\pm$ 0.0014 & 0.2751 $\pm$ 0.0043 & 3.3e-07 & 0.74 \\
\ips   & 87860               & 7.03                 & 0.0088 & 0.1764 $\pm$ 0.0014 & 0.2771 $\pm$ 0.0046 & 3.4e-07 & 1.31 \\ 
\zps   & 89917               & 7.31                 & 0.0308 & 0.1742 $\pm$ 0.0013 & 0.2531 $\pm$ 0.0048 & -3.6e-09 & 8.26 \\
	\hline
\end{tabular}
\caption{\rm{The rest-frame power law variability parameters for the complete set of {Shen} {et~al.} (2011) quasars. The "Offset" column is the mean magnitude difference between PS1 and SDSS measurements of a quasar. This amount is added to each SDSS magnitude.} }\label{tab:vrestall}
\end{table*}

Despite the added complexity of cross-matching two surveys, we see that our results in Fig. \ref{fig:f6} are very consistent with a clean power law. These power laws are in turn consistent with those measured in {Vanden Berk} {et~al.} (2004) and {Schmidt} {et~al.} (2010). Notably, we see that quasars vary more in bluer bands and have a power law index of approximately $\gamma = \Gamma/2 = 0.25$. 

Beyond confirming the basic validity of the power law model of ensemble quasar variability, the physical interpretation of these results is a bit tricky, since it mixes quasars of different redshifts and luminosities into a single result. Ultimately, the results in Table \ref{tab:vrestall} and Fig. \ref{fig:f6} are useful for a broad search of $z < 2.7$ quasars in an area not already covered by SDSS (for instance, the PS1 southern area). But they are not very useful extending our knowledge of variability into new astrophysical regimes. For that, we examine the relationship between quasar variability and the redshift, luminosity and observed wavelength of the quasars studied in Section \ref{sect:varzl}.

\section{Quasar Variability versus Redshift, Luminosity and Wavelength}\label{sect:varzl}

\begin{figure}[ht]
\includegraphics[width=0.99\columnwidth]{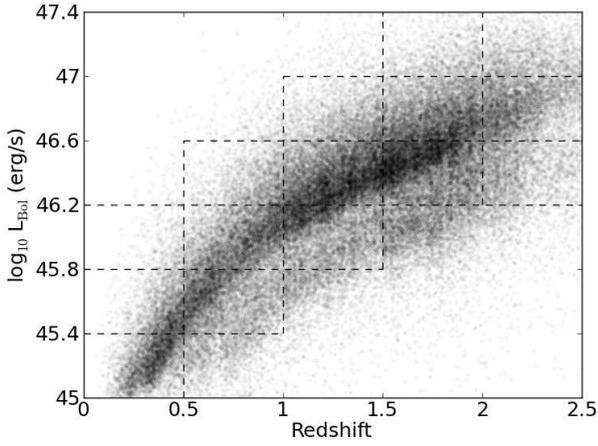}
\caption{\rm{The distribution of spectroscopically confirmed quasars from {Shen} {et~al.} (2011) in redshift-$\log \rm{L}_{\rm{Bol}}$ space. The dotted lines mark the boundaries of the regions we use in Table \ref{tab:vzlbig}.} } 
\label{fig:f12}\end{figure}

To study how quasar variability depends on redshift and luminosity, we divide our quasar sample into redshift-luminosity bins that are fairly complete and perform the fits from Section \ref{sect:varrest}. To identify these regions in redshift-luminosity space we examine Fig. \ref{fig:f12}. We bin the data in bins that are 0.5 wide in redshift and 0.4 dex wide in $\log$ L$_{\rm{Bol}}$. The centers of these bins are listed in Table \ref{tab:vzlbig}. It is difficult to evaluate the completeness of quasar catalogs in redshift-luminosity space, and the reader should note that this analysis applies to quasars in the {Shen} {et~al.} (2011) catalog (essentially $i' < 20.1$). We opt not to do this analysis with model masses, as they have even more systematic uncertainty than bolometric luminosities and are generally quite degenerate with luminosity. 

\begin{table*}
\begin{tabular}{cccccccc}
        \hline
z    & $\log $L & A$_{\gps}$        &  $\gamma_{\gps}$   & cov$_{\gps}$ & A$_{\rps}$        & $\gamma_{\rps}$   & cov$_{\rps}$ \\ 
        \hline
0.25 & 45.2 & 0.279 $\pm$ 0.012 & 0.191 $\pm$ 0.029 & -1.9e-04 & 0.227 $\pm$ 0.011 & 0.179  $\pm$ 0.028 & -1.2e-04 \\
0.25 & 45.6 & 0.246 $\pm$ 0.010 & 0.239 $\pm$ 0.026 & -8.7e-05 & 0.199 $\pm$ 0.011 & 0.258  $\pm$ 0.035 & -1.4e-04 \\
0.25 & 46.0 & 0.260 $\pm$ 0.028 & 0.150 $\pm$ 0.069 & -1.2e-03 & 0.210 $\pm$ 0.026 & 0.212  $\pm$ 0.084 & -1.4e-03 \\
0.75 & 45.6 & 0.241 $\pm$ 0.005 & 0.244 $\pm$ 0.015 & -1.5e-05 & 0.338 $\pm$ 0.009 & -0.016 $\pm$ 0.017 & -1.3e-04 \\
0.75 & 46.0 & 0.217 $\pm$ 0.005 & 0.221 $\pm$ 0.016 & -2.4e-05 & 0.199 $\pm$ 0.005 & 0.254  $\pm$ 0.016 & -1.3e-05 \\
0.75 & 46.4 & 0.183 $\pm$ 0.009 & 0.269 $\pm$ 0.034 & -7.2e-05 & 0.187 $\pm$ 0.010 & 0.185  $\pm$ 0.030 & -9.1e-05 \\
1.25 & 46.0 & 0.267 $\pm$ 0.005 & 0.223 $\pm$ 0.011 & -9.0e-07 & 0.210 $\pm$ 0.005 & 0.262  $\pm$ 0.015 & -1.7e-06 \\
1.25 & 46.4 & 0.220 $\pm$ 0.004 & 0.252 $\pm$ 0.011 & -2.1e-06 & 0.174 $\pm$ 0.004 & 0.263  $\pm$ 0.013 & -1.0e-06 \\
1.25 & 46.8 & 0.196 $\pm$ 0.008 & 0.206 $\pm$ 0.025 & -1.5e-05 & 0.165 $\pm$ 0.009 & 0.197  $\pm$ 0.037 & -6.8e-05 \\
1.75 & 46.4 & 0.250 $\pm$ 0.004 & 0.240 $\pm$ 0.010 & 2.1e-06 & 0.203  $\pm$ 0.004 & 0.284  $\pm$ 0.011 & 3.6e-06 \\
1.75 & 46.8 & 0.205 $\pm$ 0.004 & 0.248 $\pm$ 0.013 & 4.4e-06 & 0.172  $\pm$ 0.004 & 0.299  $\pm$ 0.016 & 5.7e-06 \\
1.75 & 47.2 & 0.204 $\pm$ 0.015 & 0.092 $\pm$ 0.061 & -4.6e-04 & 0.147 $\pm$ 0.012 & 0.308  $\pm$ 0.049 & 4.9e-05 \\
2.25 & 46.4 & 0.255 $\pm$ 0.008 & 0.216 $\pm$ 0.021 & 2.3e-05 & 0.226  $\pm$ 0.008 & 0.256  $\pm$ 0.021 & 3.2e-05 \\
2.25 & 46.8 & 0.212 $\pm$ 0.005 & 0.249 $\pm$ 0.017 & 1.3e-05 & 0.190  $\pm$ 0.005 & 0.287  $\pm$ 0.019 & 1.6e-05 \\
2.25 & 47.2 & 0.188 $\pm$ 0.009 & 0.243 $\pm$ 0.033 & 3.4e-05 & 0.162  $\pm$ 0.009 & 0.341  $\pm$ 0.037 & 5.1e-05 \\
	\hline
z    & $\log $L & A$_{\ips}$        & $\gamma_{\ips}$    & cov$_{\ips}$ & A$_{\zps}$        & $\gamma_{\zps}$   & cov$_{\zps}$ \\ 
        \hline
0.25 & 45.2 & 0.191 $\pm$ 0.010 & 0.199 $\pm$ 0.032 & -1.2e-04 & 0.183 $\pm$ 0.009 & 0.237 $\pm$ 0.035 & -9.8e-05 \\
0.25 & 45.6 & 0.190 $\pm$ 0.010 & 0.223 $\pm$ 0.037 & -1.5e-04 & 0.190 $\pm$ 0.010 & 0.221 $\pm$ 0.038 & -1.4e-04 \\
0.25 & 46.0 & 0.188 $\pm$ 0.025 & 0.209 $\pm$ 0.085 & -1.2e-03 & 0.200 $\pm$ 0.017 & 0.247 $\pm$ 0.067 & -3.2e-04 \\
0.75 & 45.6 & 0.209 $\pm$ 0.006 & 0.220 $\pm$ 0.018 & -2.8e-05 & 0.189 $\pm$ 0.005 & 0.242 $\pm$ 0.019 & -9.2e-06 \\
0.75 & 46.0 & 0.183 $\pm$ 0.005 & 0.294 $\pm$ 0.018 & -1.5e-05 & 0.213 $\pm$ 0.006 & 0.117 $\pm$ 0.017 & -4.5e-05 \\
0.75 & 46.4 & 0.173 $\pm$ 0.009 & 0.242 $\pm$ 0.036 & -9.5e-05 & 0.147 $\pm$ 0.008 & 0.312 $\pm$ 0.041 & -2.1e-05 \\
1.25 & 46.0 & 0.206 $\pm$ 0.005 & 0.273 $\pm$ 0.016 & -2.6e-06 & 0.202 $\pm$ 0.004 & 0.299 $\pm$ 0.017 & 3.3e-06 \\
1.25 & 46.4 & 0.166 $\pm$ 0.004 & 0.285 $\pm$ 0.013 & -1.0e-06 & 0.165 $\pm$ 0.003 & 0.328 $\pm$ 0.016 & 4.6e-06 \\
1.25 & 46.8 & 0.157 $\pm$ 0.008 & 0.193 $\pm$ 0.034 & -6.2e-05 & 0.141 $\pm$ 0.007 & 0.310 $\pm$ 0.036 & 1.5e-06 \\
1.75 & 46.4 & 0.171 $\pm$ 0.004 & 0.256 $\pm$ 0.013 & 2.8e-06 & 0.188  $\pm$ 0.004 & 0.193 $\pm$ 0.013 & -3.0e-06 \\
1.75 & 46.8 & 0.170 $\pm$ 0.005 & 0.139 $\pm$ 0.024 & -4.4e-05 & 0.145 $\pm$ 0.004 & 0.286 $\pm$ 0.019 & 9.1e-06 \\
1.75 & 47.2 & 0.155 $\pm$ 0.014 & 0.098 $\pm$ 0.054 & -1.0e-04 & 0.125 $\pm$ 0.012 & 0.248 $\pm$ 0.065 & -3.9e-05 \\
2.25 & 46.4 & 0.202 $\pm$ 0.009 & 0.181 $\pm$ 0.029 & -7.9e-06 & 0.169 $\pm$ 0.007 & 0.306 $\pm$ 0.030 & 7.3e-05 \\
2.25 & 46.8 & 0.165 $\pm$ 0.006 & 0.265 $\pm$ 0.020 & 1.4e-05 & 0.148  $\pm$ 0.005 & 0.259 $\pm$ 0.025 & 2.2e-05 \\
2.25 & 47.2 & 0.153 $\pm$ 0.009 & 0.249 $\pm$ 0.035 & 3.4e-05 & 0.125  $\pm$ 0.008 & 0.345 $\pm$ 0.048 & 1.0e-04 \\
	\hline
\end{tabular}
\caption{\rm{The fit values of A and $\gamma$ across all four bands as well as all redshift and luminosity ranges. Each row contain the results for all quasars with z within 0.25 of the "z" and $\log $L within 0.2 of "$\log $L". Note that the g and r results are in the top half of the table and the i and z results are in the bottom half. For ease, we only list the parameters used to produce the result in Table \ref{tab:vzl}} }\label{tab:vzlbig}
\end{table*}

\begin{figure}[ht]
\includegraphics[width=0.49\columnwidth]{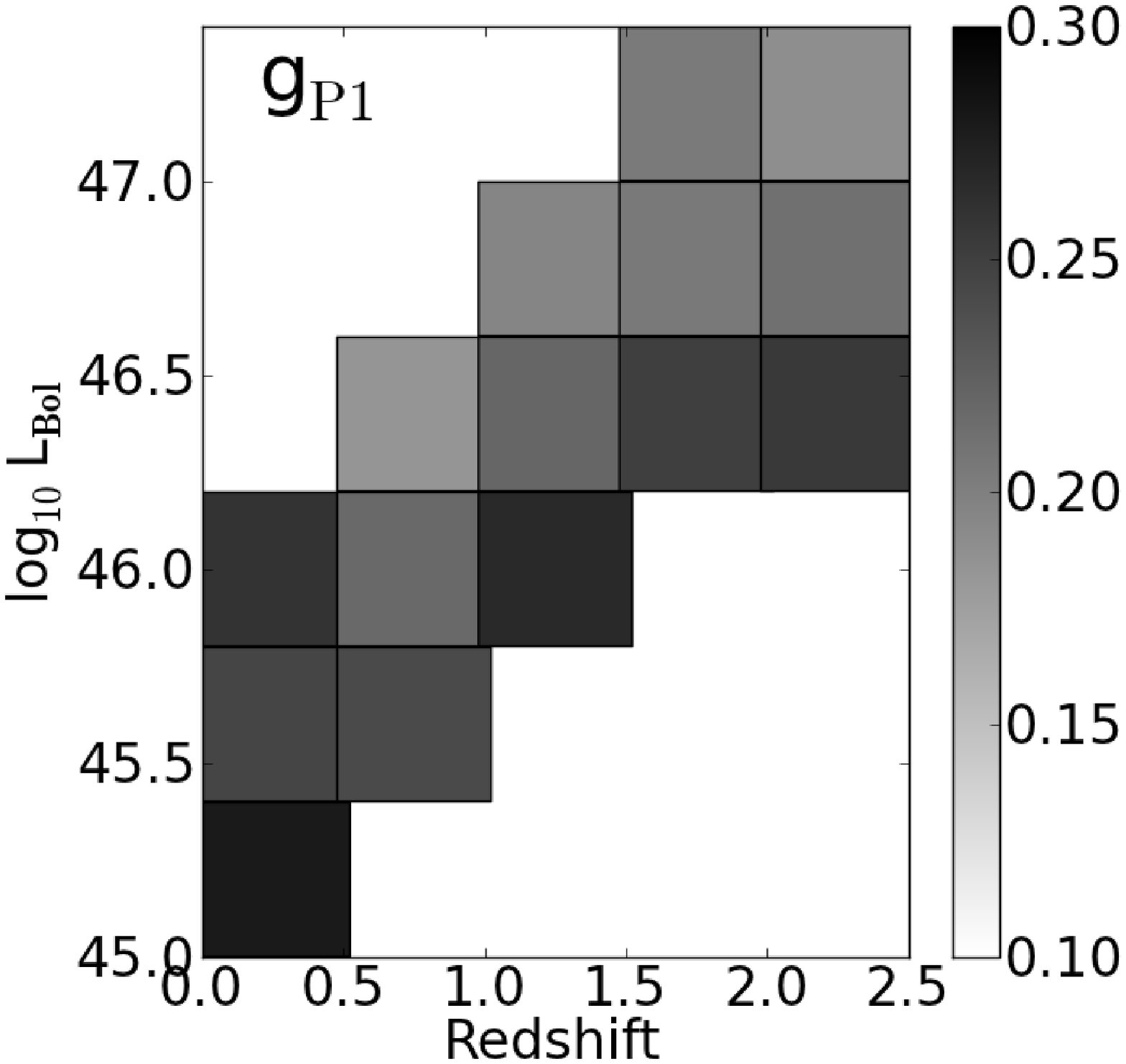}
\includegraphics[width=0.49\columnwidth]{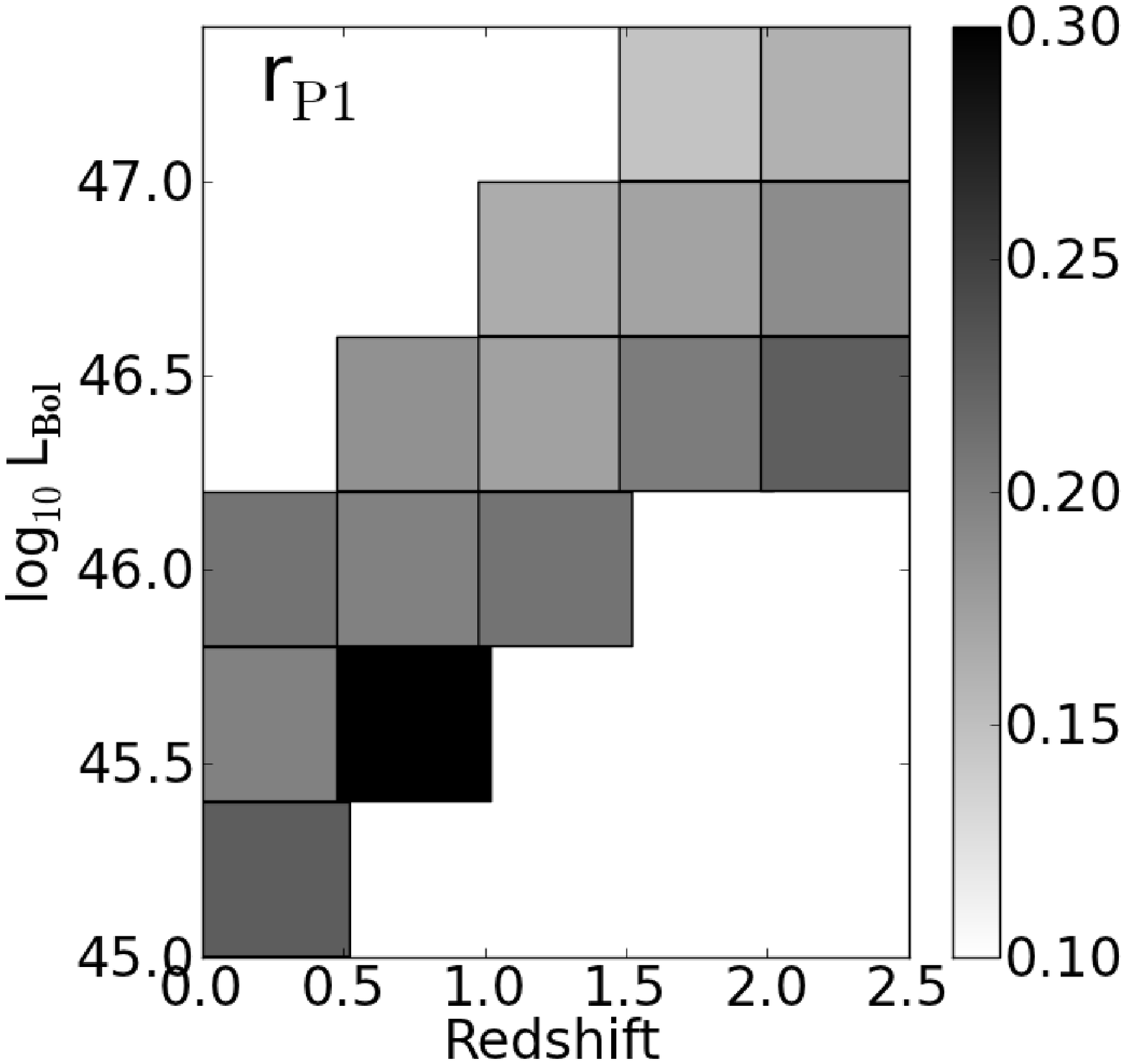}
\includegraphics[width=0.49\columnwidth]{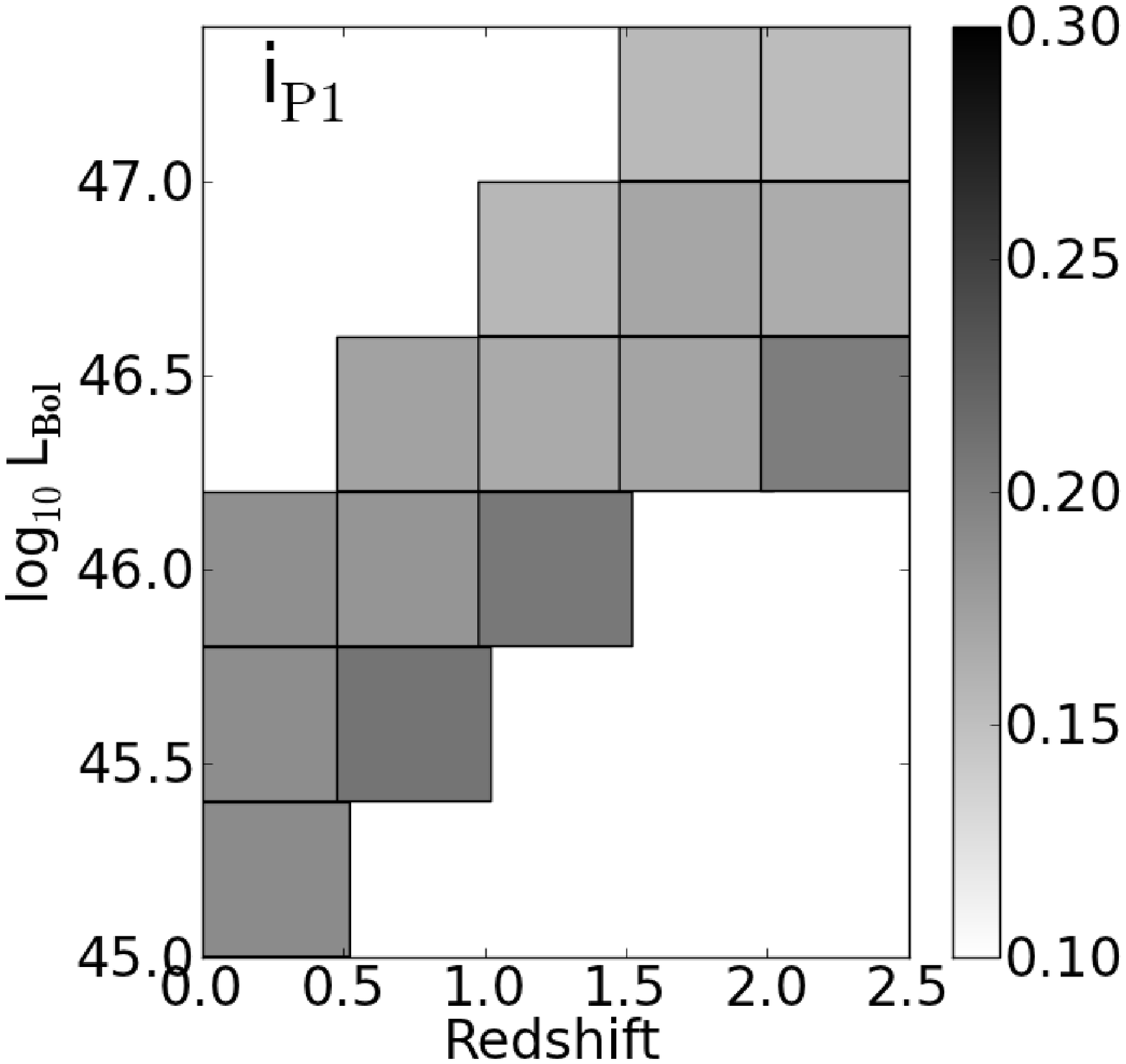}
\includegraphics[width=0.49\columnwidth]{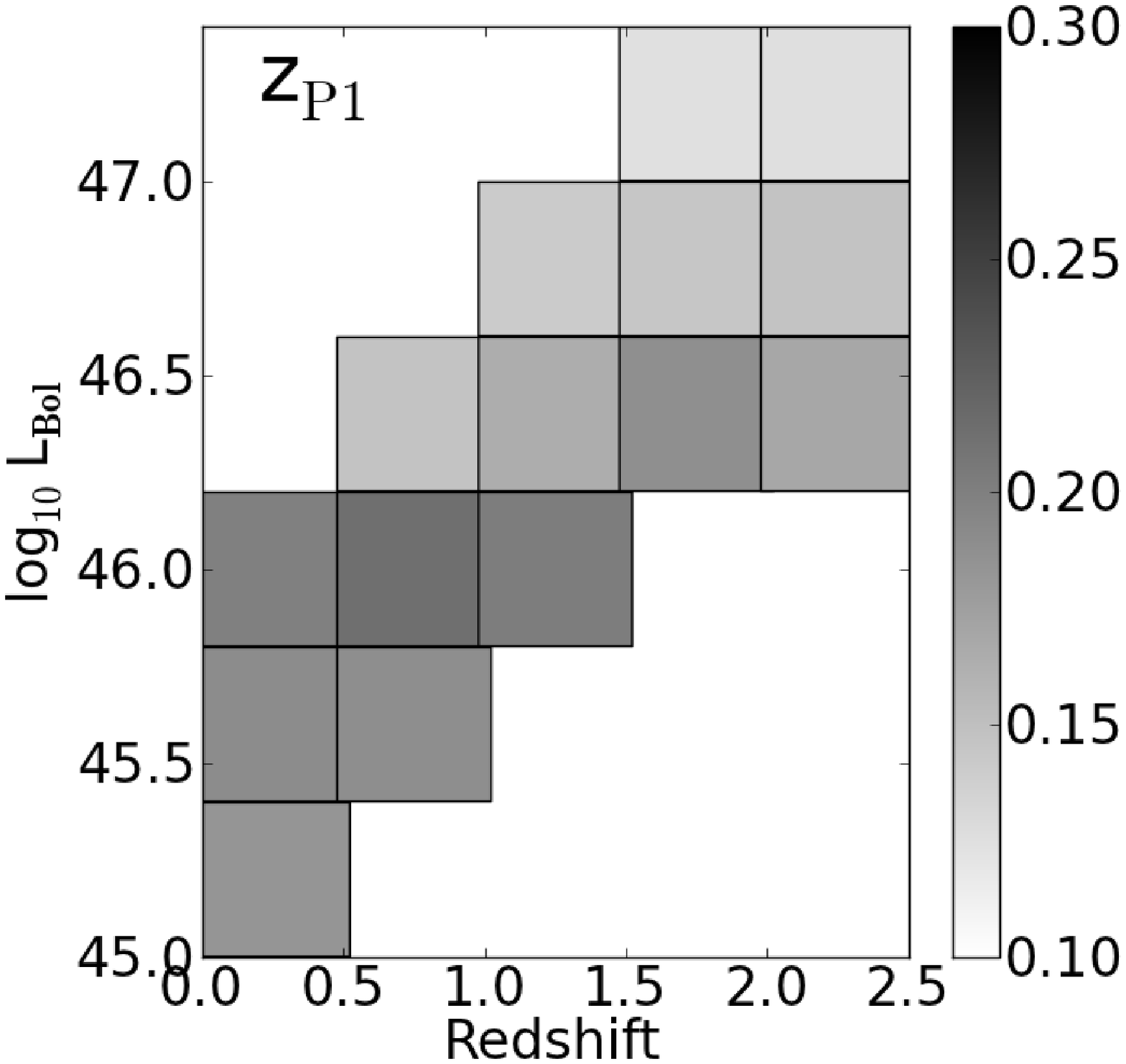}
\caption{\rm{The estimated value of A as a function of redshift and bolometric luminosity for quasars in \gps (upper left), \rps (upper right), \ips (lower left) and \zps (lower right).} } 
\label{fig:f13}\end{figure}

\begin{figure}[ht]
\includegraphics[width=0.49\columnwidth]{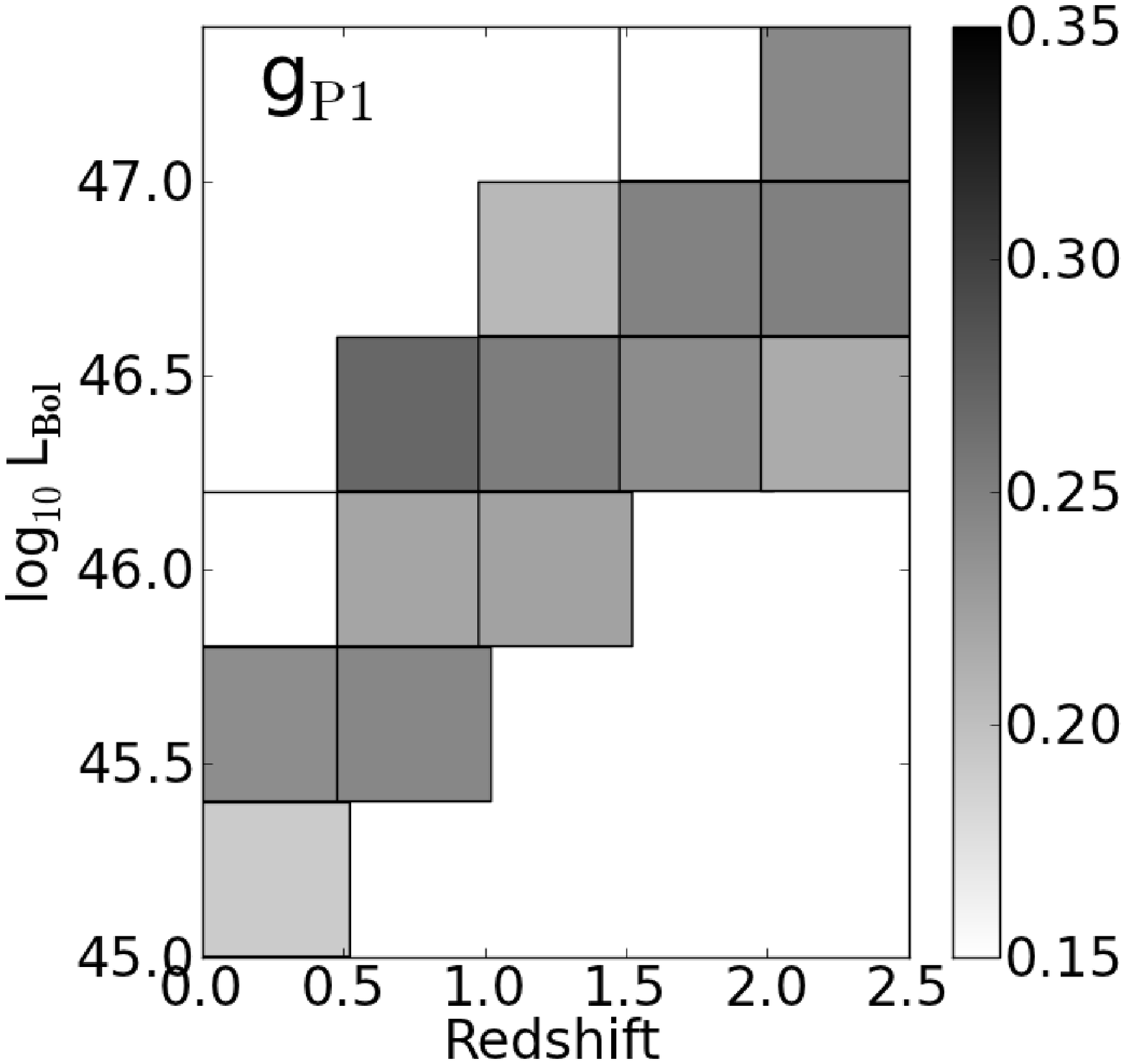}
\includegraphics[width=0.49\columnwidth]{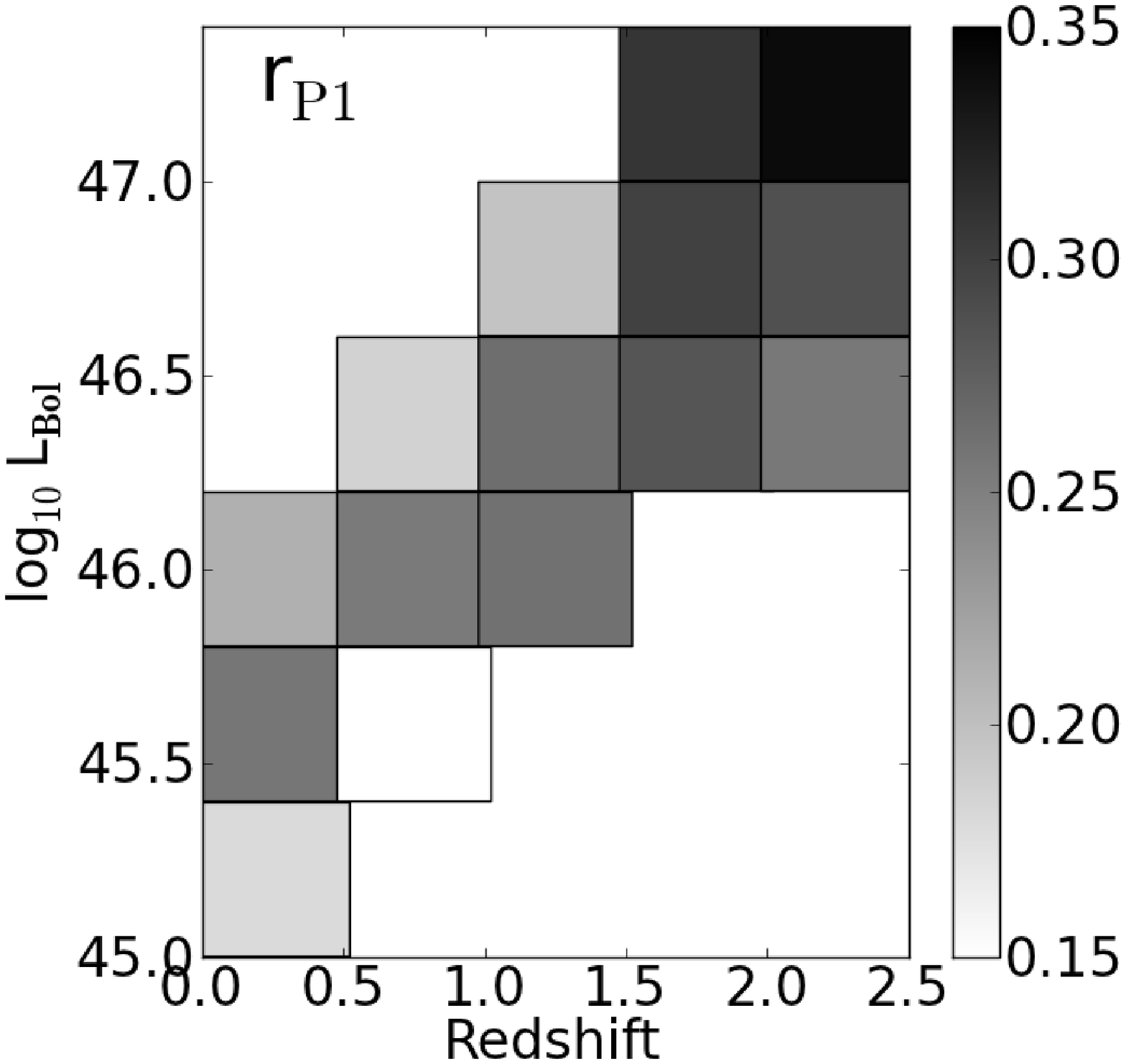}
\includegraphics[width=0.49\columnwidth]{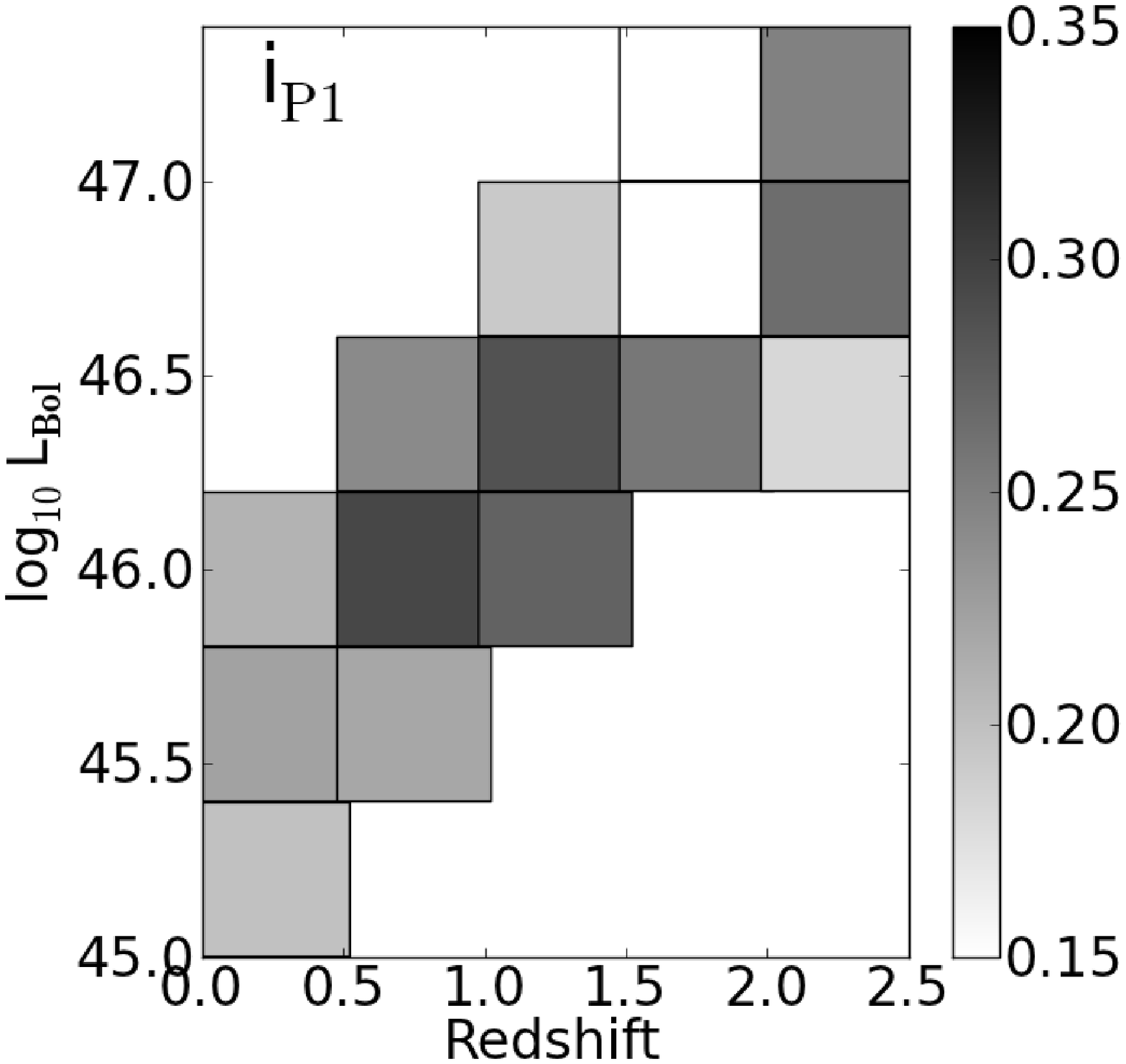}
\includegraphics[width=0.49\columnwidth]{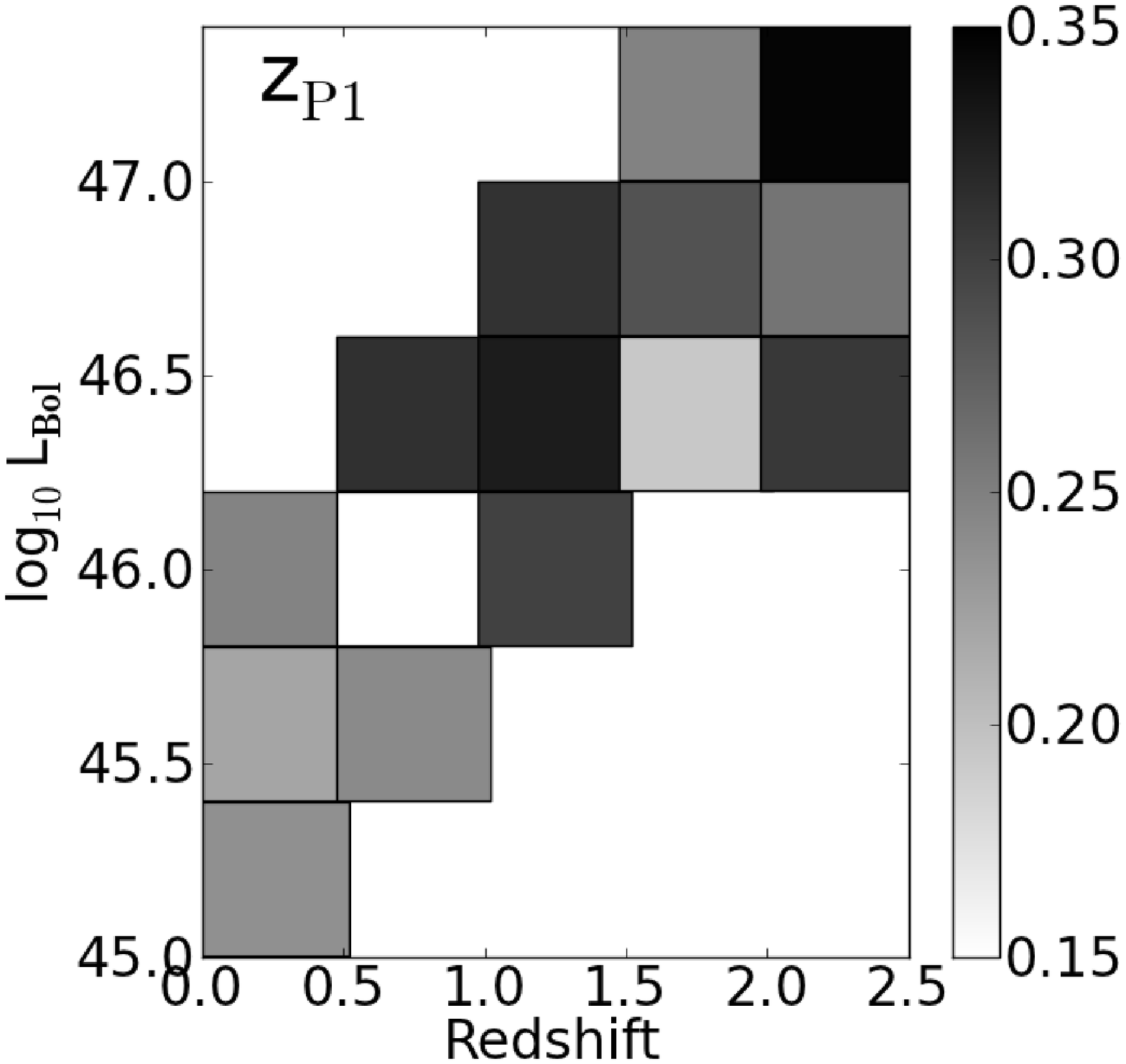}
\caption{\rm{The estimated value of $\gamma$ as a function of redshift and bolometric luminosity for quasars in \gps (upper left), \rps (upper right), \ips (lower left) and \zps (lower right).} } 
\label{fig:f14}\end{figure}

Qualitatively, we see several trends in the data in Table \ref{tab:vzlbig}, Fig. \ref{fig:f13}, and Fig. \ref{fig:f14}. The amplitude, A, tends to decrease with luminosity and increase with redshift. The power law index, $\gamma$, tends to increase with both redshift and luminosity, but in Table \ref{tab:vzlbig}, we see that the error bars on $\gamma$ are fairly large. There are several outlier points for each variable, and due to the covariance between A and $\gamma$, positive outliers in A correlate with negative outliers in $\gamma$. To see if these qualitative trends in variability are statistically robust, we parameterize variability in each filter over all redshift-luminosity space as
\begin{eqnarray}\label{eq:varzl}
\rm{A} &=& \rm{A}_0 \left(1+\rm{z}\right)^{\rm{B}_{\rm{z}}} \left(\frac{\rm{L}}{10^{46}\ \rm{erg}\ \rm{s}^{-1}}\right)^{\rm{B}_{\rm{L}}},\nonumber\\
\log \rm{A} &=& \log \rm{A}_0+\rm{B}_{\rm{z}}\log\left(1+ \rm{z}\right)+\rm{B}_{\rm{L}}\left(\log_{10} \rm{L}-46\right)\log 10,\nonumber\\
\gamma &=& \gamma_0+\beta_{\rm{z}} \rm{z}+\beta_{\rm{L}}\left(\log_{10} \rm{L}-46\right).
\end{eqnarray}
Because of the large covariance, between A and $\gamma$, we fit the two variables jointly. For ease, we use the central $\log_{10} \rm{L}_{\rm{Bol}}$ and z of each bin. 

\begin{table*}
\begin{tabular}{cccccccc}
        \hline
Filter &  A$_0$              & B$_{\rm{z}}$               & B$_{\rm{L}}$                & $\gamma_0$        & $\beta_{\rm{z}}$          & $\beta_{\rm{L}}$         & $\chi^2_{\rm{red}}$  \\
        \hline
\gps   & 0.1766 $\pm$ 0.0058 & 0.455 $\pm$ 0.045 & -0.162 $\pm$ 0.011 & 0.236 $\pm$ 0.016 & -0.003 $\pm$ 0.013 & 0.016 $\pm$ 0.017 & 2.98\\
\rps   & 0.1597 $\pm$ 0.0062 & 0.445 $\pm$ 0.053 & -0.213 $\pm$ 0.013 & 0.210 $\pm$ 0.017 & 0.010 $\pm$ 0.015 & 0.090 $\pm$ 0.018 & 9.38\\
\ips   & 0.1577 $\pm$ 0.0068 & 0.248 $\pm$ 0.060 & -0.125 $\pm$ 0.015 & 0.258 $\pm$ 0.020 & -0.006 $\pm$ 0.017 & -0.002 $\pm$ 0.022 & 3.58\\
\zps   & 0.1570 $\pm$ 0.0062 & 0.234 $\pm$ 0.055 & -0.148 $\pm$ 0.014 & 0.243 $\pm$ 0.020 & -0.004 $\pm$ 0.018 & 0.060 $\pm$ 0.023 & 7.91\\
	\hline
\end{tabular}
\caption{\rm{The fitted values from the variables in Eq. \ref{eq:varzl} for each filter. Because of the significant covariance between A and $\gamma$, the two variables are fit simultaneously and there is only one $\chi^2_{\rm{red}}$. Note that A$_0$ and $\gamma_0$ are not mean values, but the fitted values at L=$10^{46}$, z=0.} }\label{tab:vzl}
\end{table*}

The results of the four fits are in Table \ref{tab:vzl}. Note that A$_0$ and $\gamma_0$ are the values of these variables at L=$10^{46}$, z=0, not some mean estimate. In every case, we see that variability amplitude does indeed increase with redshift and decrease with luminosity. Both these trends make intuitive sense. Higher luminosity quasars also tend to be more massive and larger. So whatever physical process drives an individual quasar's variability, we would expect brighter quasars to vary more slowly. Quasars are also generally more variable in bluer wavelengths. For a given filter, higher redshift quasars are being observed at bluer rest-frame wavelengths. This may explain the redshift dependence. We also see that variation in $\gamma$ is essentially consistent with 0. We refit the data with a single B$_{\rm{z}}$, B$_{\rm{L}}$ and $\gamma_0$ to obtain 
\begin{eqnarray}\label{eq:varzl2} 
\rm{A} &=& \rm{A}_0 \left(1+\rm{z}\right)^{B_{\rm{z}}} \left(\frac{\rm{L}}{10^{46} \rm{erg}\ \rm{s}^{-1}}\right)^{B_{\rm{L}}},\nonumber\\
\rm{A}_{0\rm{g}} &=& 0.1897 \pm 0.0037,\nonumber\\
\rm{A}_{0\rm{r}} &=& 0.1620 \pm 0.0032,\nonumber\\
\rm{A}_{0\rm{i}} &=& 0.1465 \pm 0.0030,\nonumber\\
\rm{A}_{0\rm{z}} &=& 0.1416 \pm 0.0028,\nonumber\\
\gamma &=& 0.2457 \pm 0.0025,\nonumber\\
B_{\rm{z}} &=& 0.365 \pm 0.026,\nonumber\\
B_{\rm{L}} &=& -0.159 \pm 0.006.
\end{eqnarray} 
For completeness, we note that $\chi^2 = 6.29$. It is a better fit than the two worst single filter fits. These results are consistent with, but more precise than those found in {Vanden Berk} {et~al.} (2004). 

The above results are useful for quasars in PS1 or similar surveys, but to generalize, we can also account for wavelength dependency. For each quasar and in each filter, we calculate a variability amplitude assuming a constant $\gamma = 0.2457$. We also calculate a rest frame wavelength, $\lambda$, defined as the central wavelength of the $\griz$ filters (483 nm, 619 nm, 752 nm and 866 nm, respectively) divided by $1+z$. We then bin our quasars by $\lambda$ in five evenly spaced bins between 1500 nm and 6500 nm and use the same L and z bins from Table \ref{tab:vzlbig}. In each bin, we take the median as our amplitude measurement and the difference between the 10th and 90th percentile divided by the square root of the number of sources as our error bars. We also ignore bins with fewer than 30 light curves. These last steps ensure robustness. Finally, we fit the 3D binned data with Eq. \ref{eq:varzl2} with an additional $\lambda$ term to obtain:
 
%We can then fit A$_0(\lambda)$ as a power law to obtain
%\begin{eqnarray} 
%A_0 &=& A_{1000} \left(\frac{\lambda}{1000\ \rm{nm}}\right)^\nu\nonumber\\
%A_{1000}&=&0.1289 \pm 0.0027,\ \nu = -0.512 \pm 0.045
%\end{eqnarray} 
%with a reduced $\chi^2 = 1.53$. It would be difficult to map this relationship to a physical quasar model, since quasar emission cannot be modeled by a simple power law. But the fact that variability's dependence on redshift is smaller than its dependence on observed $\lambda$ suggests that redshift dependence may indeed be entirely due to observing light that is bluer in the quasar's rest-frame. 

\begin{eqnarray}\label{eq:varzl3}
\rm{A} &=& \rm{A}_0 \left(1+\rm{z}\right)^{B_{\rm{z}}} \left(\frac{\rm{L}}{10^{46} \rm{erg}\ \rm{s}^{-1}}\right)^{B_{\rm{L}}} \left(\frac{\lambda}{1000\ \rm{nm}}\right)^\nu ,\nonumber\\
\rm{A}_0&=&0.0789 \pm 0.0017,\nonumber\\ 
B_{\rm{z}} &=& 0.153 \pm 0.028,\nonumber\\
B_{\rm{L}} &=& -0.200 \pm 0.006,\nonumber\\
\nu &=& -0.441 \pm 0.018.
\end{eqnarray}
For completeness, we note that $\chi^2 = 113$ with 78 degrees of freedom. But our error bars are non-canonical.

The change in variability amplitude versus luminosity is surprisingly small in both Eq. \ref{eq:varzl2} and Eq. \ref{eq:varzl3}. It suggests that a "bright" quasar, 100 times more luminous than a "dim" quasar varies with roughly half the amplitude. If one assumes that observed quasar luminosity scale roughly with mass, this defies simple dimensional arguments based on the Schwarzschild radius scaling as M. The increase of variability with redshift is consistent with the idea that galaxies were generally more active and dynamic in the early universe. The inverse relationship between variability and $\lambda$ is also generally accepted ({MacLeod} {et~al.} 2012). 

\begin{figure}[ht]
\includegraphics[width=0.99\columnwidth]{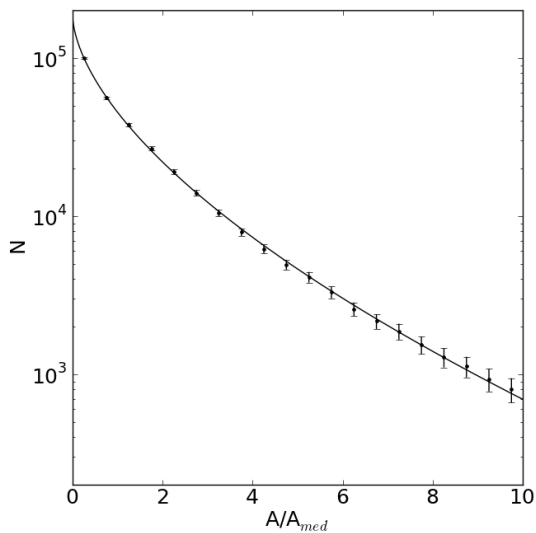}
\caption{\rm{The distribution of actual variability amplitudes over the average amplitudes defined in Eq. \ref{eq:varzl3}.} } 
\label{fig:f15}\end{figure}
We anticipate that these results will be useful for estimating quasar variability in theoretical models and simulations. With that in mind, we make a simple fit for the distribution of variability amplitudes of all quasars with respect to the average amplitude defined in Eq. \ref{eq:varzl3}. For each light curve (subscripted "i"), we take the ratio of variability amplitude over the expected amplitude, A$_i$/A and bin the results in Fig. \ref{fig:f15}. We find that globally, individual variability amplitudes are distributed as:
\begin{eqnarray}\label{eq:vardist}
P(x_i) &=& 1.16 e^{-1.4 x_i^{0.6}},\nonumber\\
x_i&=&\frac{\rm{A}_i}{\rm{A}}.
\end{eqnarray}
Again, A is defined for all z, L and $\lambda$ in Eq. \ref{eq:varzl3}, and we have assumed a constant exponent of $\gamma = 0.2457$. This model of quasar variability is obviously simplistic, but may nonetheless be useful for simulations and models across a wide range of z, L and $\lambda$.

\section{Conclusions}

The cross-matched PS1-SDSS catalog we produced and studied here is a powerful tool for probing quasar variability. The large sample size of $10^5$ quasars and the 10 year time spanned by SDSS and PS1 allowed us to study ensemble quasar variability with unprecedented precision. When we examined the typical quasar root mean squared magnitude variability as parameterized by the structure function, V(t) = A t$^{\gamma}$, we found ${\gamma} \approx 0.25$ regardless of quasar luminosity, quasar redshift or observer filter. We confirmed the well known trend that quasars are more variable in bluer bands than in redder bands. 
%We also found that the analytic DRW model of quasar variability is a sightly better model for parameterizing our ensemble structure model than the power law model, but we continue to use the power law because of computational speed in large surveys and the simplicity of combining information from power law fits from different filters.

After measuring observer-frame quasar variability, we examined the effectiveness of using quasar variability as a method for quasar selection with our current sample. We produced a simple variability cut that rejected 92\% of unresolved objects with quasar-like colors while recovering 71\% of known spectroscopic quasars. If we restrict our search to SDSS-WISE objects that can be fairly cleanly separated into quasars and non-quasars, we find that our cut produces a sample that is 67\% complete and 48\% pure. Using only $griz$ color cuts, our sample is only 4.1\% pure. PS1-SDSS quasar variability selection will become more powerful as more PS1 epochs are taken and added to our database. 

Another interesting application of the current dataset was studying the more astrophysical problem of rest-frame quasar variability. As in the observer-frame case, the structure function with V(t) = A t$^{\gamma}$ where ${\gamma} \approx 0.25$ worked very well in all cases. By dividing quasars into redshift-luminosity bins, we were able to show that the 1 year amplitude of variation, A, scales with redshift, z, as $(1+z)^{0.365}$ and with luminosity, L, as $L^{-0.159}$. The redshift dependence is partly due to the fact that for a given filter in the observer-frame, we are studying the rest-frame bluer part of a quasar's spectrum when the quasar is at higher redshift. Finally, we fit variability with respect to redshift, luminosity and wavelength across all four filters as $\rm{V}(\rm{z},\rm{L},\lambda,\rm{t}) = 0.079 (1+\rm{z})^{0.15}(\rm{L}/\rm{L}_0)^{-0.2} (\lambda/1000\ nm)^{-0.44} (\rm{t}/1 yr)^{0.246}$ where $\rm{L}_0 = 10^{46} \rm{erg} \rm{s}^{-1}$ . 

In the future, we will extend this work to at least three new research areas. First, searching for quasars with variability should help us find many of the $z > 2.5$ quasars that are missed by u' band selection. Secondly, we can use the above techniques to find high redshift lensed quasars. All the database level cross-matching and variability measuring in this paper has been performed for aperture magnitudes as well as the PSF magnitudes we discussed here. Looking for extended objects which vary in their aperture magnitudes with a quasar-like signature is a promising new way to find quasar lenses. Finally, while cross-matching the {Shen} {et~al.} (2011) quasars with PS1, we have identified many quasars that changed by extreme amounts (sometimes more than two magnitudes). Studying the properties of these highly variable quasars and being able to have spectra both before and after their large change should provide a unique look into quasar variability and cataclysmic events.  

\section{Acknowledgments}

The PS1 Surveys have been made possible through contributions of the Institute for Astronomy, the University of Hawaii, the Pan-STARRS Project Office, the Max-Planck Society, and its participating institutes, the Max Planck Institute for Astronomy, Heidelberg, and the Max Planck Institute for Extraterrestrial Physics, Garching, The Johns Hopkins University, Durham University, the University of Edinburgh, Queen's University Belfast, the Harvard-Smithsonian Center for Astrophysics, and the Las Cumbres Observatory Global Telescope Network, Incorporated, the National Central University of Taiwan, and the National Aeronautics and Space Administration under Grant No. NNX08AR22G issued through the Planetary Science Division of the NASA Science Mission Directorate. 
%We also thank Dr. Chelsea MacLeod for data used in section \ref{sect:drw}.

%\appendix

\section{Appendix A: Parameterizing Quasar Variability for Measurement Pairs with Approximately Constant Time Intervals}\label{sect:selection}

We want to estimate the variability amplitude, A, of a quasar implied by a set of measurement pairs that produces a set of magnitude differences, m$_i$, time differences, t$_i$ and statistical uncertainties, $\sigma_i$. We assume the time intervals are roughly constant (that we are binning measurement pairs into fairly fine time bins). We start with the likelihood in Eq. \ref{eq:logp} and differentiate it with respect to A$^2$ to find the probability maximum
\begin{equation}
\frac{\partial\log(\rm{P})}{\partial (\rm{A}^2)} = -\frac{1}{2} \sum_i \frac{\rm{t}_i^\Gamma}{\rm{A}^2\rm{t}_i^\Gamma+\sigma_i^2} - \frac{\rm{m}_i^2 \rm{t}_i^\Gamma}{\left(\rm{A}^2\rm{t}_i^\Gamma+\sigma_i^2\right)^2}=0.
\end{equation}
In the case that the variation of $\sigma_i^2$ and t$_i^\gamma$ are totally negligible, this is trivially solvable as
\begin{eqnarray}\label{eq:A0}
\rm{A}_0^2=\frac{\rm{m}_0^2-\sigma_0^2}{\rm{t}_0^\Gamma},\ \rm{m}_0^2=\frac{1}{\rm{n}}\sum_i \rm{m}_i^2,\nonumber\\
\sigma^2_0=\frac{1}{\rm{n}}\sum_i \sigma_i^2,\ \rm{t}_0^\Gamma=\frac{1}{\rm{n}}\sum_i \rm{t}_i^\Gamma.\label{eq:approx} 
\end{eqnarray}
Formally, A$_0^2$ can be less than 0 for sources that randomly vary less than their error bars suggest. We allow this to happen for individual time bins for individual quasars to avoid biasing our results. Ultimately, A$^2$ is positive for every time bin in the in every ensemble average we produce. 

We can improve the result in Eq. \ref{eq:approx} by allowing for variations from A$_0^2$, $\sigma^2_0$ and t$_0^\Gamma$ and treating them as perturbations. This produces
\begin{eqnarray}
\frac{\partial\log(\rm{P})}{\partial (\rm{A}^2)} = -\frac{1}{2} \sum_i \left(\frac{\rm{t}_0^\Gamma+\delta \rm{t}_i^\Gamma}{\rm{m}_0^2+\rm{t}_0^\Gamma \delta\rm{A}^2+\rm{A}_0^2\delta \rm{t}_i^\Gamma+\delta \sigma^2_i}\right.\nonumber\\
\left. - \frac{\rm{m}_i^2\left(\rm{t}_0^\Gamma+\delta\rm{t}_i^\Gamma\right)}{\left(\rm{m}_0^2+\rm{t}_0^\Gamma\delta\rm{A}^2+\rm{A}_0^2\delta \rm{t}_i^\Gamma+\delta\sigma^2_i\right)^2}\right)=0,\nonumber\\
\delta \rm{A}^2 = \rm{A}^2-\rm{A}_0^2,\ \delta\sigma^2_i=\sigma_i^2-\sigma_0^2,\ \delta \rm{t}_i^\Gamma= \rm{t}_i^\Gamma-\delta \rm{t}_0^\Gamma.
\end{eqnarray}

We convert our terms to unit-free versions to simplify our equation and obtain 
\begin{eqnarray}
\frac{-\rm{t}_0^\Gamma}{2\rm{m}_0^2}&&\sum_i \frac{1+\tau_{1i}}{1+\rm{a}+\tau_{2i}+\rm{s}_i} - \frac{\mu_i\left(1+\tau_{1i}\right)}{\left(1+\rm{a}+\tau_{2i}+\rm{s}_i\right)^2}=0,\nonumber\\
\rm{a}&=&\frac{\rm{t}_0^\Gamma \delta \rm{A}^2}{\rm{m}_0^2},\ \mu_i=\frac{\rm{m}_i^2}{\rm{m}_0^2},\ \rm{s}_i=\frac{\delta \sigma^2_i}{\rm{m}^2_0},\nonumber\\
 \tau_{1i}&=&\frac{\delta \rm{t}_i^\Gamma}{\rm{t}_0^\Gamma},\ \tau_{2i}=\frac{\rm{A}_0^2\delta \rm{t}_i^\Gamma}{\rm{m}_0^2}.
\end{eqnarray}

A first order perturbation yields 
\begin{eqnarray}
\sum_i \left(1+\tau_{1i}-\rm{a}-\tau_{2i}-\rm{s}_i\right.&&\nonumber\\
\left. - \mu_i-\mu_i\tau_{1i}+2\mu_i\rm{a}+2\mu_i\tau_{2i}+2\mu_i\rm{s}_i\right)&=&0,\\
\sum_i \rm{a}-\mu_i\tau_{1i}+2\mu_i\tau_{2i}+2\mu_i\rm{s}_i&=&0,\\
\rm{a}=\frac{1}{\rm{n}} \sum_i \mu_i\tau_{1i}-2\mu_i\tau_{2i}-2 \mu_i\rm{s}_i.&&
\end{eqnarray}
We use the facts that $\sum_i \mu_i =\ $n$\ = \sum_i 1$ and that  $\sum_i \rm{s}_i = \sum_i \tau_{1i} = \sum_i \tau_{2i} = 0$ in the simplifications above. 

We substitute more physical variables to obtain
\begin{eqnarray}
\rm{a}&=&\frac{1}{\rm{n}\ \rm{m}_0^2}\sum_i \rm{m}_i^2\left(\left(\frac{\rm{t}_i^\Gamma-\rm{t}_0^\Gamma}{\rm{t}_0^\Gamma} \right)\right.\nonumber\\
&&\left.-2\left(\frac{\rm{A}_0^2\left(\rm{t}_i^\Gamma -\rm{t}_0^\Gamma\right)}{\rm{m}_0^2}\right)-2\left(\frac{\sigma^2_i-\sigma^2_0}{\rm{m}^2_0}\right)\right),\\
\rm{a}&=&\frac{1}{\rm{m}_0^2}\left(\left(2 \rm{A}_0^2 \rm{t}_0^\Gamma-\rm{m}_0^2\right)\left(1-\frac{1}{\rm{n}} \sum_i \frac{\rm{m}_i^2 \rm{t}_i^\Gamma}{\rm{m}_0^2 \rm{t}_0^\Gamma}\right)\right.\nonumber\\ 
&&\left.+2 \sigma_0^2-\frac{2}{\rm{n}} \sum_i \frac{\rm{m}_i^2 \sigma^2_i}{\rm{m}^2_0}\right),\\
\rm{a}&=&1+\frac{1}{\rm{m}_0^2}\left(\frac{2 \sigma_0^2}{\rm{n}} \sum_i \frac{\rm{m}_i^2 \rm{t}_i^\Gamma}{\rm{m}_0^2 \rm{t}_0^\Gamma}-\frac{\rm{m}_i^2 \rm{t}_i^\Gamma}{\rm{t}_0^\Gamma}  -\frac{2}{\rm{n}} \sum_i \frac{\rm{m}_i^2 \sigma^2_i}{\rm{m}^2_0}\right).
\end{eqnarray}
Which we add as a perturbation to obtain A$^2$:
\begin{eqnarray}
\rm{A}^2&=&\rm{A}_0^2+\frac{\rm{m}_0^2}{\rm{t}_0^\Gamma}\rm{a}\\
\rm{A}^2&=&\frac{2\rm{m}_0^2-\sigma_0^2}{\rm{t}_0^\Gamma}-\frac{1}{\rm{n}\ \rm{t}_0^\Gamma}\sum_i \frac{\rm{m}_i^2 \rm{t}_i^\Gamma}{\rm{t}_0^\Gamma} \nonumber\\
&&+\frac{2}{\rm{n}\ \rm{m}_0^2 \rm{t}_0^\Gamma}\left(\sigma_0^2 \sum_i \frac{\rm{m}_i^2 \rm{t}_i^\Gamma}{\rm{t}_0^\Gamma}-\sum_i \rm{m}_i^2 \sigma^2_i\right).\label{eq:formallycorrect}
\end{eqnarray}

While Eq. \ref{eq:formallycorrect}  is formally correct and perfectly usable for objects with many observations, objects which have been observed only a few times can have m$_0^2$ go to zero. Having m$_0^2$ in the denominator can produce arbitrarily large variability estimates. To avoid this problem, we replace the denominator m$_0^2$ with a model  
\begin{equation}
\rm{m}_0^2=\frac{1}{\rm{n}}\sum_i \rm{A}^2_{\rm{model}} \rm{t}_i^\Gamma + \sigma_i^2=\rm{A}^2_{\rm{model}}\rm{t}_0^\Gamma+\sigma_0^2
\end{equation}
and obtain a more robust solution
\begin{eqnarray}
\rm{A}^2&=&\frac{2\rm{m}_0^2-\sigma_0^2}{\rm{t}_0^\Gamma}-\frac{1}{\rm{n}\ \rm{t}_0^\Gamma}\sum_i \frac{\rm{m}_i^2 \rm{t}_i^\Gamma}{\rm{t}_0^\Gamma}\nonumber\\ 
&&+\frac{2}{\rm{n}\ \rm{t}_0^\Gamma\left(\rm{A}^2_{\rm{model}}\rm{t}_0^\Gamma+\sigma_0^2\right)}\left(\sigma_0^2 \sum_i \frac{\rm{m}_i^2 \rm{t}_i^\Gamma}{\rm{t}_0^\Gamma}-\sum_i \rm{m}_i^2 \sigma^2_i\right).\label{eq:a}
\end{eqnarray}

In the case where the time intervals are exactly constant or the variability has no apparent time dependence (for instance, a variable star that varies on time scales shorter than those that are measured), V(t) is just a constant V, $\Gamma = 0$, and Eq. \ref{eq:a} can be simplified to produce
\begin{equation}
\rm{V}^2=\rm{m}_0^2-\sigma_0^2+\frac{2}{\rm{V}^2_{\rm{model}}+\sigma_0^2}\left(\sigma_0^2 \rm{m}_0^2-\frac{1}{\rm{n}}\sum_i \rm{m}_i^2 \sigma^2_i\right).
\end{equation}

If error bars are also constant, we obtain
\begin{equation}
\rm{V}^2=\rm{m}_0^2-\sigma_0^2.
\end{equation}

We can derive a weight (inverse variance) for A$^2$ by assuming it is Gaussian distributed. The weight is then just
\begin{eqnarray}
\rm{weight}&=&-\frac{\partial^2\log(\rm{P})}{\partial (A^2)^2},\\ 
\rm{weight}&=& \frac{-1}{2} \sum_i \frac{\rm{t}_i^{\Gamma2}}{\left(\rm{V}^2(\rm{t}_i)+\sigma_i^2\right)^2} - \frac{2\rm{m}_i^2 \rm{t}_i^{\Gamma2}}{\left(\rm{V}^2(\rm{t}_i)+\sigma_i^2\right)^3},\\
\rm{weight}&=&\frac{1}{2} \sum_i \frac{\rm{t}_i^{\Gamma2}}{\left(\rm{V}^2(\rm{t}_i)+\sigma_i^2\right)^3}\left(2\rm{m}_i^2 -\rm{V}^2(\rm{t}_i)-\sigma_i^2\right).\label{eq:weight}
\end{eqnarray}
Here V is necessarily a model V and not derived from the current m$_0^2$. If one uses the V$_i$ = m$_i^2-\sigma_i^2$ as in Eq. \ref{eq:A0}, the weight will be inversely proportional to m$_0^4$. This effectively downweights all large variability data points and severely biases the final result downward. In fact, even in Eq. \ref{eq:weight}, $\sum_i \rm{m}_i^2$ will randomly be very small and can lead to negative weights. It is best to replace $\rm{m}_i^2$ with the model value V$^2$(t$_i$)-$\sigma_i^2$ and obtain
\begin{eqnarray}
\rm{weight}=\frac{1}{2} \sum_i \frac{\rm{t}_i^{\Gamma2}}{\left(\rm{V}^2(\rm{t}_i)+\sigma_i^2\right)^2},\\
\rm{weight}=\frac{1}{2} \sum_i \frac{1}{\left(\rm{A}^2_{\rm{model}}+\sigma_i^2/\rm{t}_i^\Gamma\right)^2}.\label{eq:weight2}
\end{eqnarray}
Since the weight can be solved for directly, no further approximations are necessary.

%We make the same substitutions and Taylor expansion as above to obtain
%\begin{eqnarray}
%\rm{weight} = \frac{-\rm{t}_0^{\gamma2}}{2 \rm{m}_0^4} \sum_i 1+2\tau_{1i}-2\rm{a}-2\rm{s}_i-2\tau_{2i}-2\mu_i-4\mu_i \tau_{1i}+6\mu_i a+6\mu_i \rm{s}_i+6\mu_i\tau_{2i}\\
%\rm{weight} = \frac{\rm{t}_0^{\gamma2}}{2 \rm{m}_0^4} \sum_i 1-4\rm{a}-6\mu_i \rm{s}_i+4\mu_i \tau_{1i}-6\mu_i\tau_{2i} \\
%\rm{weight} = \frac{\rm{t}_0^{\gamma2}}{2 \rm{m}_0^4} \sum_i 1-4\rm{a}+\mu_i \tau_{1i}+3\left(\mu_i \tau_{1i}-2\mu_i \rm{s}_i-2\mu_i\tau_{2i}\right), \\
%\rm{weight} = \rm{t}_0^\gamma\frac{\rm{t}_0^\gamma\rm{n}\left(1-\rm{a}\right)+\sum_i \mu_i \left(\rm{t}_i^\gamma-\rm{t}_0^\gamma \right)}{2 \rm{m}_0^4},\\ 
%\rm{weight} = \rm{t}_0^\gamma\frac{-\rm{t}_0^\gamma\rm{n}\ \rm{a}+\sum_i \frac{\rm{m}_i^2 \rm{t}_i^\gamma}{\rm{m}_0^2}}{2 \rm{m}_0^4},\\ 
%\rm{weight} = \rm{n}\rm{t}_0^\gamma \frac{-\rm{t}_0^\gamma +2\frac{\rm{t}_0^\gamma}{\rm{n}\ \rm{m}_0^2} \sum_i \frac{\rm{m}_i^2 \sigma^2_i}{\rm{m}^2_0}+\frac{2}{\rm{n}\ \rm{m}_0^2}\left(\rm{m}_0^2-\sigma_0^2\right) \sum_i \frac{\rm{m}_i^2 \rm{t}_i^\gamma}{\rm{m}_0^2}  }{2 \rm{m}_0^4} 
%\end{eqnarray}
The simple weight in Eq. \ref{eq:weight2} is proportional to the number of time intervals being examined and makes the faulty assumption that all time intervals in the set t$_i$ are independent. This is not generally true, and many time intervals will overlap and should be statistically downweighted. If we have j time intervals t$_0$ long over the time interval $\Delta$t which runs from the first measurement to the last measurement then on average, $e^{-j \rm{t}_0/\Delta t}$ of the larger interval $\Delta$t is not "covered" by at least one of the j measurements. The j+1 measurement should be weighted by approximately that factor. Thus, instead of being weighted by a factor "n", our measurements should be weighted by a factor
\begin{equation}
\rm{\hat{n}} = \sum_i e^{-i \frac{\rm{t}_0}{\Delta \rm{t}}} = \frac{1-\left(e^{-\rm{t}_0/\Delta \rm{t}}\right)^{\rm{n}}}{ 1-e^{-\rm{t}_0/\Delta \rm{t}}},
\end{equation}
and our actual weight is 
\begin{equation}
\rm{weight}=\frac{\rm{\hat{n}}}{2\rm{n}} \sum_i \frac{1}{\left(\rm{A}^2_{\rm{model}}+\sigma_i^2/\rm{t}_i^\Gamma\right)^2}.
\end{equation}

In the quick sinusoidally varying case, this reduces to
\begin{eqnarray}
\rm{weight}_{\rm{V}}=\frac{\rm{\hat{n}}}{2\rm{n}} \sum_i \frac{1}{\left(\rm{V}^2_{\rm{model}}+\sigma_i^2\right)^2}.\\
\nonumber
\end{eqnarray}

If we also impose constant error bars, we obtain
\begin{equation}
\rm{weight}_{\rm{V}}=\frac{\rm{\hat{n}}}{2\left(\rm{V}^2_{\rm{model}}+\sigma_0^2\right)^2}.
\end{equation}

%% \bibliography


\begin{references}


































































%%
%% End of file `sample.tex'.
\reference {Abazajian}	 {Abazajian}, K.~N., {Adelman-McCarthy}, J.~K., {Ag{\"u}eros}, M.~A., {et al.} 2009, \apjs, 182, 543
\reference {Aihara}	 {Aihara}, H., {Allende Prieto}, C., {An}, D., {et al.} 2011, \apjs, 193, 29
\reference {Antonucci}	 {Antonucci}, R. 1993, \araa, 31, 473
\reference {Brada{\v c}}	 {Brada{\v c}}, M., {Schneider}, P., {Steinmetz}, M., {et al.} 2002, \aap, 388, 373
\reference {Butler}	 {Butler}, N.~R. \& {Bloom}, J.~S. 2011, \aj, 141, 93
\reference {Chambers}	 {Chambers}, K.~C. 2011, in Bulletin of the American Astronomical Society,  Vol.~43, American Astronomical Society Meeting Abstracts 217, 222.02--+
\reference {Dalal}	 {Dalal}, N. \& {Kochanek}, C.~S. 2002, \apj, 572, 25
\reference {Dobler}	 {Dobler}, G. \& {Keeton}, C.~R. 2006, \mnras, 365, 1243
\reference {Eyer}	 {Eyer}, L. 2002, AcA, 52, 241
\reference {Fan}	 {Fan}, X. 1999, \aj, 117, 2528
\reference {Fan}	 {Fan}, X., {Narayanan}, V.~K., {Lupton}, R.~H., {et al.} 2001, \aj, 122, 2833
\reference {Geha}	 {Geha}, M., {Alcock}, C., {Allsman}, R.~A., {et al.} 2003, \aj, 125, 1
\reference {Giveon}	 {Giveon}, U., {Maoz}, D., {Kaspi}, S., {et al.} 1999,  \mnras, 306, 637
\reference {Hopkins}	 {Hopkins}, P.~F., {Hernquist}, L., {Cox}, T.~J., {et al.} 2006, \apjs, 163, 1
\reference {Inada}	 {Inada}, N., {Oguri}, M., {Becker}, R.~H., {et al.} 2008, \aj, 135, 496
\reference {Juric}	 {Juric}, M. 2011, in Bulletin of the American Astronomical Society, Vol.~43,  American Astronomical Society Meeting Abstracts 217, 433.19
\reference {Kaiser}	 {Kaiser}, N., {Aussel}, H., {Burke}, B.~E., {et al.} 2002, in Society of Photo-Optical  Instrumentation Engineers (SPIE) Conference Series, Vol. 4836, Society of  Photo-Optical Instrumentation Engineers (SPIE) Conference Series, ed.  {J.~A.~Tyson \& S.~Wolff}, 154--164
\reference {Kaiser}	 {Kaiser}, N., {Burgett}, W., {Chambers}, K., {et al.} 2010,  in Society of Photo-Optical Instrumentation Engineers (SPIE) Conference  Series, Vol. 7733, Society of Photo-Optical Instrumentation Engineers (SPIE)  Conference Series
\reference {Kawaguchi}	 {Kawaguchi}, T., {Mineshige}, S., {Umemura}, M., {et al.} 1998, \apj,  504, 671
\reference {Kelly}	 {Kelly}, B.~C., {Bechtold}, J., \& {Siemiginowska}, A. 2009, \apj, 698, 895
\reference {Kembhavi}	 {Kembhavi}, A.~K. \& {Narlikar}, J.~V. 1999, {Quasars and active galactic  nuclei : an introduction} ({Cambridge University Press})
\reference {Kim}	 {Kim}, D.-W., {Protopapas}, P., {Byun}, Y.-I., {et al.} 2011, \apj, 735, 68
\reference {Koz{\l}owski}	 {Koz{\l}owski}, S., {Kochanek}, C.~S., {Jacyszyn}, A.~M., {et al.} 2012, \apj, 746, 27
\reference {Koz{\l}owski}	 {Koz{\l}owski}, S., {Onken}, C.~A.,  {Kochanek}, C.~S., {et al.} 2013, \apj, 775, 92
\reference {Kozlowski}	 {Kozlowski}, S., {Kochanek}, C.~S., \& {Udalski}, A. 2011, VizieR Online Data  Catalog, 219, 40022
\reference {Koz{\l}owski}	 {Koz{\l}owski}, S., {Kochanek}, C.~S., {Udalski}, A., {et al.} 2010, \apj, 708, 927
\reference {Lovegrove}	 {Lovegrove}, J., {Schild}, R.~E., \& {Leiter}, D. 2011, \mnras, 412, 2631
\reference {MacLeod}	 {MacLeod}, C.~L., {Brooks}, K., {Ivezi{\'c}}, {\v Z}., {et al.} 2011, \apj, 728, 26
\reference {MacLeod}	 {MacLeod}, C.~L., {Ivezi{\'c}}, {\v Z}., {Kochanek}, C.~S., {et al.} 2010, \apj,  721, 1014
\reference {MacLeod}	 {MacLeod}, C.~L., {Ivezi{\'c}}, {\v Z}., {Sesar}, B., {et al.} 2012,  \apj, 753, 106
\reference {Morgan}	 {Morgan}, C.~W., {Kochanek}, C.~S., {Morgan}, N.~D., {et al.} 2010,  \apj, 712, 1129
\reference {Morganson}	 {Morganson}, E., {De Rosa}, G., {Decarli}, R., {et al.} 2012, \aj, 143, 142
\reference {Mortlock}	 {Mortlock}, D.~J., {Warren}, S.~J., {Venemans}, B.~P., {et al.} 2011,  \nat, 474, 616
\reference {Mushotzky}	 {Mushotzky}, R.~F., {Edelson}, R., {Baumgartner}, W., {et al.} 2011,  \apjl, 743, L12
\reference {Oguri}	 {Oguri}, M., {Inada}, N., {Pindor}, B., {et al.} 2006, \aj, 132, 999
\reference {Oguri}	 {Oguri}, M. \& {Marshall}, P.~J. 2010, \mnras, 405, 2579
\reference {Osmer}	 {Osmer}, P.~S. 1982, \apj, 253, 28
\reference {Palanque-Delabrouille}	 {Palanque-Delabrouille}, N., {Yeche}, C., {Myers}, A.~D., {et al.} 2011, \aap, 530, A122
\reference {Pereyra}	 {Pereyra}, N.~A., {Vanden Berk}, D.~E., {Turnshek}, D.~A., {et al.} 2006,  \apj, 642, 87
\reference {Rees}	 {Rees}, M.~J. 1984, \araa, 22, 471
\reference {Richards}	 {Richards}, G.~T., {Fan}, X., {Newberg}, H.~J., {et al.} 2002, \aj, 123, 2945
\reference {Richards}	 {Richards}, G.~T., {Myers}, A.~D., {Gray}, A.~G., {et al.} 2009, \apjs, 180, 67
\reference {Richards}	 {Richards}, G.~T., {Nichol}, R.~C., {Gray}, A.~G., {et al.} 2004, \apjs, 155, 257
\reference {Riechers}	 {Riechers}, D.~A., {Walter}, F., {Carilli}, C.~L., {et al.} 2007a, \apjl, 671, L13
\reference {Riechers}	 {Riechers}, D.~A., {Walter}, F., {Cox}, P., {et al.} 2007b, \apj, 666, 778
\reference {Schlafly}	 {Schlafly}, E.~F., {Finkbeiner}, D.~P., {Juric}, M., {et al.} 2012, ArXiv e-prints
\reference {Schmidt}	 {Schmidt}, K.~B., {Marshall}, P.~J., {Rix}, H.-W., {et al.} 2010, \apj, 714, 1194
\reference {Schneider}	 {Schneider}, D.~P., {Richards}, G.~T., {Fan}, X., {et al.} 2002, \aj,  123, 567
\reference {Schneider}	 {Schneider}, D.~P., {Richards}, G.~T., {Hall}, P.~B., {et al.} 2010, \aj, 139, 2360
\reference {Sesar}	 {Sesar}, B., {Ivezi{\'c}}, {\v Z}., {Grammer}, S.~H., {et al.} 2010, \apj, 708, 717
\reference {Shen}	 {Shen}, Y., {Richards}, G.~T., {Strauss}, M.~A., {et al.} 2011, \apjs,  194, 45
\reference {Tonry}	 {Tonry}, J.~L., {Stubbs}, C.~W., {Lykke}, K.~R., {et al.} 2012, \apj, 750, 99
\reference {Vanden Berk}	 {Vanden Berk}, D.~E., {Wilhite}, B.~C., {Kron}, R.~G., {et al.} 2004, \apj, 601, 692
\reference {Villata}	 {Villata}, M., {Raiteri}, C.~M., {Balonek}, T.~J., {et al.} 2006, \aap, 453, 817
\reference {Wambsganss}	 {Wambsganss}, J. 2006, in Saas-Fee Advanced Course 33: Gravitational Lensing:  Strong, Weak and Micro, ed. G.~{Meylan}, P.~{Jetzer}, P.~{North},  P.~{Schneider}, C.~S. {Kochanek}, \& J.~{Wambsganss}, 453--540
\reference {Warren}	 {Warren}, S.~J., {Hewett}, P.~C., {Irwin}, M.~J., {et al.} 1991,  \apjs, 76, 1
\reference {Wright}	 {Wright}, E.~L., {Eisenhardt}, P.~R.~M., {Mainzer}, A.~K., {et al.} 2010, \aj, 140, 1868
\reference {Wu}	 {Wu}, X.-B., {Hao}, G., {Jia}, Z., {et al.} 2012, \aj, 144,  49
\reference {York}	 {York}, D.~G., {Adelman}, J., {Anderson}, Jr., {et al.} 2000, \aj, 120, 1579
\reference {Yun}	 {Yun}, M.~S., {Scoville}, N.~Z., {Carrasco}, J.~J., {et al.} 1997,  \apjl, 479, L9
\reference {Zu}	 {Zu}, Y., {Kochanek}, C.~S., {Koz{\l}owski}, S., {et al.} 2012, \apj, 765, 106
\end{references}
\end{document}